\documentclass[pra,nobalancelastpage,twocolumn,nofootinbib,superscriptaddress]{revtex4-2}
\pdfoutput=1

\usepackage{hyperref}
\usepackage{color,amsthm,amsmath,amsxtra,amsfonts,dsfont,graphicx,bm,amssymb}
\usepackage{natbib}
\usepackage[dvipsnames]{xcolor}
\usepackage{bm}
\usepackage{subcaption}
\def\Id{{\openone}}
\newcommand{\be}{\begin{equation}}
\newcommand{\ee}{\end{equation}}
\newcommand{\bea}{\begin{eqnarray}}
\newcommand{\eea}{\end{eqnarray}}
\newcommand{\bse}{\begin{subequations}}
\newcommand{\ese}{\end{subequations}}

\newcommand{\ket}[1]{\vert#1\rangle}
\newcommand{\bra}[1]{\langle#1\vert}
\newcommand{\tr}{\mathrm{tr}}

\newcommand{\hvar}[0]{\bar{h}}

\newcommand{\randsigma}[0]{s}
\newcommand{\paulisigma}[0]{\sigma}

\begin{document}
\title{Preparation and verification of tensor network states}
\author{Esther Cruz}
\affiliation{Max-Planck-Institute of Quantum Optics,
Hans-Kopfermann-Stra\ss{}e 1, 85748 Garching, Germany, and\\
Munich Center for Quantum Science and Technology,
Schellingstra\ss{}e~4, 80799 M\"unchen, Germany}
\author{Flavio Baccari}
\affiliation{Max-Planck-Institute of Quantum Optics,
Hans-Kopfermann-Stra\ss{}e 1, 85748 Garching, Germany, and\\
Munich Center for Quantum Science and Technology,
Schellingstra\ss{}e~4, 80799 M\"unchen, Germany}
\author{Jordi Tura}
\affiliation{Max-Planck-Institute of Quantum Optics,
Hans-Kopfermann-Stra\ss{}e 1, 85748 Garching, Germany, and\\
Munich Center for Quantum Science and Technology,
Schellingstra\ss{}e~4, 80799 M\"unchen, Germany}
\affiliation{Instituut-Lorentz, Universiteit Leiden, P.O. Box 9506, 2300 RA Leiden, The Netherlands}
\author{Norbert Schuch}
\affiliation{Max-Planck-Institute of Quantum Optics,
Hans-Kopfermann-Stra\ss{}e 1, 85748 Garching, Germany, and\\
Munich Center for Quantum Science and Technology,
Schellingstra\ss{}e~4, 80799 M\"unchen, Germany}
\affiliation{University of Vienna, Faculty of Mathematics, Oskar-Morgenstern-Platz 1, 1090 Wien, Austria, and\\
University of Vienna, Faculty of Physics, Boltzmanngasse
5, 1090 Wien, Austria
}
\author{J. Ignacio Cirac}
\affiliation{Max-Planck-Institute of Quantum Optics,
Hans-Kopfermann-Stra\ss{}e 1, 85748 Garching, Germany, and\\
Munich Center for Quantum Science and Technology,
Schellingstra\ss{}e~4, 80799 M\"unchen, Germany}

\begin{abstract}
We consider a family of tensor network states defined on regular lattices that come with a natural definition of an adiabatic path to prepare them. This family comprises relevant classes of states, such as injective Matrix Product and Projected Entangled-Pair States, and some corresponding to classical spin models. We show how uniform lower bounds to the gap of the parent Hamiltonian along the adiabatic trajectory can be efficiently computed using semi-definite programming. This allows one to check whether the adiabatic preparation can be performed efficiently with a scalable effort. We also derive a set of observables whose expectation values can be easily determined and that form a complete set, in the sense that they uniquely characterize the state. We identify a subset of those observables which can be efficiently computed if one has access to the quantum state and local measurements, and analyze how they can be used in verification procedures. 
\end{abstract}

\maketitle
% ================================================================
\section{Introduction}

The simulation of many-body states is one of the most promising and long-awaited applications of quantum computing. In particular, quantum computers are expected to efficiently prepare certain entangled multipartite states, which can help us in the study of quantum many-body systems, or in variational quantum algorithms. The advent of the first generations of both analog and quantum computers has triggered a strong interest in the preparation of such states. For instance, GHZ states up to tens of qubits have been prepared with trapped ions \cite{sackett_2000, Leibfried_2005, Monz_2011, Friis_2018, pogorelov2021compact}, Rydberg atoms \cite{Omran_2019}, superconducting qubits \cite{DiCarlo_2010, Song_2017, Gong_2019, Song_2019}, photons \cite{Bouwmeester_1999, Pan_2001, Zhong_2018, Wang_2018} or nuclear spins \cite{laflamme1998nmr, neumann2008multipartite}. 

Tensor network states (TNS) constitute an especially appealing family of multipartite states \cite{cirac2020matrix}. On the one hand, they are expected to efficiently approximate the ground states of local Hamiltonians. On the other, many paradigmatic states in the realm of quantum information or condensed matter physics are simple examples of TNS. The best-known class of such states is that of Matrix Product States (MPS) \cite{Werner}, which corresponds to a one-dimensional geometry. Higher-dimensional generalizations are known as Projected Entangled Pair States (PEPS) \cite{Verstraete_2004}. In both cases, they are characterized by the bond dimension, $D$, which is directly related to the size of the tensors building such states. Those states possess a special property, namely that they are the ground state of a local, frustration-free Hamiltonian. This implies that, in case of no degeneracy, one can easily check the successful preparation of the state by measuring a set of local observables. Thus, such states naturally play an essential role in the certification of quantum computers \cite{universalblind, reichardt2012classical, reichard2013nature, Wiebe_2014, Wiebe_2014_2, Wiebe_2015, aharonov2017interactive, fitzsimons2015post, Hangleiter_2017, mahadev2018classical, brakerski2019cryptographic, aaronson2016complexitytheoretic, Boixo_2018}.

The preparation of TNS has been actively pursued in the last years and, in particular, methods that operate efficiently in terms of the number of qudits (or lattice size), $N$, have been devised.\footnote{By efficient we mean that the computational time grows at most polynomially with $N$. We will also use the short cut ``exponential time" meaning that the time scales exponentially with $N$.} Matrix Product States can be sequentially generated \cite{SchoenSequentallyGen} in a time that scales linearly with $N$. In fact, MPS of up to ten qubits have been recently prepared in a superconducting setup \cite{Besse_2020, Schwartz434, Istrati_2020, Takedaeaaw4530}. Certain kinds of PEPS (the so-called sequentially generated) can also be prepared in the same time scale \cite{Ba_uls_2008} and proposals for the generation of sequentially generated PEPS have been recently put forward \cite{Pichler_2017, Gimeno_Segovia_2019}. While containing many paradigmatic examples of TNS, those states are fine-tuned in the sense that a small change on the tensors defining the state may lead to another PEPS outside that class. In fact, those tensors are strongly restricted by the fact that the states have to be sequentially generated.

In \cite{Ge_2016}, a very efficient quantum algorithm to generate a wide range of PEPS was introduced. That class of states is stable under deformations of the tensors and thus they are not fine-tuned. The algorithm is based on the adiabatic method and the circuit depth scales as $O(\log(N))$. The algorithm needs, however, the existence of a gap, $\Delta$, along the adiabatic path, something which is difficult to ensure since checking that typically requires a computational time that scales exponentially with $N$. Additionally, it is devised for digital quantum computers, but not analog ones.

In this paper we consider a family of states on arbitrary lattices and prove two novel results on this family. First, we show how the computation of the gap of the parent Hamiltonian can be expressed as a semidefinite programming problem (SDP), and how this allows us to efficiently compute lower bounds $\delta\le \Delta$ on the gap. Second, we show that for such families of states, it is possible to predict the expectation values of many observables beyond those appearing as terms of the parent Hamiltonian. In fact, there is an exponential number of such observables, which forms a complete set in the set of operators acting on the many-body Hilbert space. 

The first result naturally leads to preparation protocols by using for instance the one in Ref. \cite{Ge_2016}, since we can efficiently identify the subset of tensor network states for which a bound in the gap is known. Besides, we also extend the adiabatic algorithm to continuous time, which is more suitable for analog quantum computers.

The second result naturally leads to certification protocols based on interactive proofs (see \cite{universalblind, reichardt2012classical, reichard2013nature, Wiebe_2014, Wiebe_2014_2, Wiebe_2015,  aharonov2017interactive, mahadev2018classical, brakerski2019cryptographic, aaronson2016complexitytheoretic, Boixo_2018} for previous works on interactive verification schemes) which are inspired by the difficulty of sampling from PEPS \cite{Schuch_2007, Haferkamp_2020}. While they require a certification time that grows exponentially with $N$ \cite{aaronson2016complexitytheoretic, Boixo_2018}, we propose efficient versions that, however, rely on stronger standard 
complexity assumptions. We also explain that, in case they can be spoofed, this would immediately lead to new  classical algorithms to estimate physically relevant expectation values of observables in the class of states we consider. 

This paper is structured as follows. In Section \ref{section_states} we present the class of states and their corresponding parent Hamiltonian. The states depend on two positive parameters, $t, \beta \geq 0$, that can be viewed as time and inverse temperature, respectively. We show that, by construction, these states can be smoothly connected to a product state. 
This family of states is esentially the same as the one considered in  Ref. \cite{Ge_2016}, although our formulation allows to extend the adiabatic quantum algorithm of \cite{Ge_2016} to continuous time in a natural fashion. We will see later how this family includes injective MPS and PEPS, as well as some ground states of classical models whose PEPS description might not be injective (although their parent Hamiltonian has unique ground state).
In Section \ref{section_GapsCorr} we first show how one can efficiently find lower bounds on the gap of the parent Hamiltonian by means of an SDP. In particular, for every value of $t$ we can find a maximum value of $\beta(t)$ such that for all $\beta < \beta(t)$ the gap of the Hamiltonian can be lower bounded by a constant that does not depend on the system size. We then introduce sets of operators whose expectation values can be easily computed and that are complete, in the sense that they provide a tomography of the state. Finally, we propose verification protocols in Section \ref{section_verification} and discuss some possible complexity arguments.

% ================================================================
\section{States and parent Hamiltonian}\label{section_states}

In this section, we introduce the family of states that we will consider in the present work, which is comparable to the one considered in Ref. \cite{Ge_2016}. It is built in terms of sets of local commuting operators, together with two parameters $t,\beta\ge 0$. For $t=\beta=0$, they are product states, whereas otherwise they are entangled and can be efficiently expressed as PEPS. Based on that fact, we will explicitly construct a frustration-free parent Hamiltonian, which will play an important role in the procedure to prepare the states. 
We then analyse how to prepare these family of states adiabatically and study the scaling of the computational time as a function of the system size and a lower bound on the gap. We build upon the work done in Ref. \cite{Ge_2016} and extend their adiabatic algorithm to continuous times. We review here the main idea of the algorithm from Ref. \cite{Ge_2016} and its runtime, and refer to the reader to the original paper for a more technical discussion. Note that while in Ref. \cite{Ge_2016} a constant gap was assumed, we will present later in Section \ref{section_GapsCorr} a method for lower-bounding such gap, which constitutes our main result regarding ground state preparation. This will immediately allow us to know which states we can efficiently prepare. Lastly, we show how the presented family of states includes many physically relevant examples of TN states.
% -------------------------------------------------------------
\subsection{Setting}

We will consider a rather general setup, although we will give examples later for regular lattices. We consider $N$ qudits, with Hilbert space ${\cal H}_d=\mathds{C}^d$ and computational basis $\{|0\rangle,\ldots,|d-1\rangle\}$, located at the vertices ${\cal V}$ of a graph ${\cal G}({\cal V},{\cal E})$ with edges ${\cal E}$ of bounded degree $z$ (that is, the maximal number of edges starting from a given vertex). We define ${\cal H}={\cal H}_d^{\otimes N}$, and denote the set of Hermitian operators acting on ${\cal H}$
by ${\cal A}$. For any $O\in {\cal A}$, we define its support $\lambda(O)\subset{\cal V}$ as the subset of vertices such that $O=\mathrm{tr}_i (O) \otimes \Id_i/d$ if and only if $i\notin \lambda(O)$, where  $\mathrm{tr}_i$ is the trace with respect to the qudit at vertex $i$.

The graph $\mathcal G$ defines a natural distance $d(i,j)$ between two vertices $i,j\in {\cal V}$ as the minimal number of edges connecting them. We also define the radius of a subset of vertices $\lambda\subset {\cal V}$,
 \be
 r(\lambda) = \min_{i\in {\cal \lambda}} \max_{j\in{\cal \lambda}} d(i,j).
 \ee
 We denote by ${\cal A}_r\in {\cal A}$ the set of Hermitian operators acting on $\mathcal H$ whose support has a radius of at most $r$.

% -------------------------------------------------------------
\subsection{States}

We call $\mathbb{K}_{r,M}\subset {\cal A}$ the set of operators that can be written as 
 \be
 \label{decomposition}
 K = \sum_{n=1}^M \kappa_n \, , \quad [\kappa_n,\kappa_m]=0 \, \, , \,  n,m=1,\ldots,M  \ ,
 \ee
where $\kappa_n\in {\cal A}_r$, $\|\kappa_n\|_{\infty}\le 1$ and $\kappa_n \succeq 0$. That is, it is the set that can be written as a sum of $M$ commuting, positive semidefinite, subnormalized operators that have a radius of at most $r$.
Apart from the trivial cases, where the operators $\kappa_n$ act on single vertices or when they are products of the same single-qudit operator, one can easily construct non-trivial sets $\mathbb{K}_{r,M}$. In Appendix \ref{appendix_operators} we briefly review some of them.

Given $r_{\alpha},M_{\alpha}\in \mathbb{N}$, $K_{\alpha}\in \mathbb{K}_{r_{\alpha},M_{\alpha}}$, for $\alpha=1,2$, we define the family of states
 \be
 \label{Psibetat}
 \lvert\Psi(\beta,t)\rangle = \frac{1}{Z(\beta,t)} e^{\beta K_1} e^{i t K_2} \lvert\varphi_1 \rangle \otimes \ldots \otimes \lvert \varphi_N\rangle\ ,
 \ee
where $\beta,t\ge 0$, $Z$ is a normalization constant and $\ket{\varphi_i}$ are arbitrary single-qudit states. This family of states obviously contains all product states if we take $\beta= t = 0$. For $\beta=0$, we have $Z(0,t)=1$. Note that we only explicitly denote the dependence of $\ket{\Psi(\beta,t)}$ on $\beta$ and $t$, while omitting the dependence on $K_{\alpha}$, for $\alpha=1,2$, and $\ket{\varphi_i}$, in order to ease the notation.

\begin{figure*}
    \centering
    \includegraphics[width=0.80\linewidth]{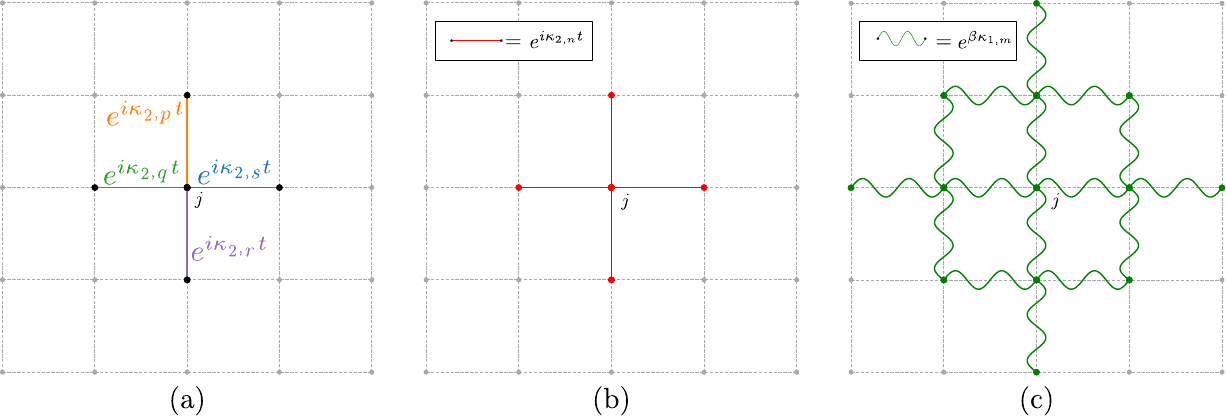}
    \caption{Example of operators $e^{i\kappa_{2,n}t}$ (straight lines) and $e^{\beta\kappa_{1,m}}$ (wiggled lines) that act each on two adjacent sites.  (a) All operators $e^{i\kappa_{2,n}t}$ that act on site $j$. Here $\mu_j = \{ p,q,r,s \}$. (b) Support of the operator $ \prod_{n\in \mu_j} e^{-i \kappa_{2,n} t}$. (c)  Support of the operator $\prod_{m\in \nu_j} e^{-\beta \kappa_{1,m}}$. }
    \label{fig:vertices_operators}
\end{figure*}

% -------------------------------------------------------------
\subsection{Parent Hamiltonian} \label{section_parent}
We now show that any state \eqref{Psibetat} is the unique ground states of a local, frustration-free Hamiltonian, which we construct explicitly. We denote by $\kappa_{\alpha,n}$ the operators appearing in the decomposition \eqref{decomposition} of $K_{\alpha}$, for $\alpha = 1,2$. For $j\in \mathcal V$, let $\mu_j:=\{n\,|\,j\in\lambda(\kappa_{2,n})\}$ be the index set of all terms $\kappa_{2,n}$ which act non-trivially on site $j$, and $\nu_j=\{m\,|\,\exists\, n\in\mu_j:\,\lambda(\kappa_{1,m})\cap\lambda(\kappa_{2,n})\ne\emptyset\}$ the index set of all terms $\kappa_{1,m}$ which overlap with one of the previous terms. See Figure \ref{fig:vertices_operators} for an example of how such set of vertices would be.
Then, for $j=1,\dots,N$, we define
 \be
 \label{hj}
 h_j = O_j^\dagger \Pi_j O_j\ ,
 \ee
where $\Pi_{j} = \mathds{1}_j - | \varphi_j  \rangle\langle \varphi_j|_j$ acts on the qudit at vertex $j$, and
 \be
 \label{eq:def-Oj}
 O_j = \prod_{n\in \mu_j} e^{-i \kappa_{2,n} t} \prod_{m\in \nu_j} e^{-\beta \kappa_{1,m}}\ ,
 \ee
which is invertible. With this definition, we have $h_j=h_j^\dagger \succeq 0$ and $h_j |\Psi(\beta,t)\rangle=0$.

We define the parent Hamiltonian of $\ket{\Psi(\beta,t)}$ as
 \be
 \label{eq:def-Hbetat}
 H(\beta,t) = \sum_{j=1}^N h_j\ .
 \ee
Note that we have now also suppressed the dependence of $h_j$ on $\beta$ and $t$ for convenience.

Let us show that, indeed, \eqref{Psibetat} is the unique ground state of such an operator. Since $h_j\ket{\Psi(\beta,t)}=0$, we have that 
$H(\beta,t) |\Psi(\beta,t)\rangle=0$, and since 
$H(\beta,t)\succeq0$, this implies that $\ket{\Psi(\beta,t)}$ is a ground state of $H(\beta,t)$, with ground state energy $0$.
Conversely, if $H(\beta,t)|\Psi'\rangle=0$, then $h_j\ket{\Psi'}=0$ and thus $\Pi_j e^{-i t K_2} e^{-\beta K_1} |\Psi'\rangle=0$ for all $j$, which in turn means that $\ket{\Psi'}$ is proportional to  $\ket{\Psi(\beta,t)}$: We thus see that $\ket{\Psi(\beta,t)}$ is the  unique ground state of $H(\beta,t)$. 

% -------------------------------------------------------------

\subsection{Adiabatic preparation}\label{section_Adiabatic}

The existence of a smooth path of Hamiltonians connecting $H(\beta,t)$, and thus $\ket{\Psi(\beta,t)}$, to a simple product state at $H(0,0)$ implies that these states can be prepared adiabatically, by starting with the product state $\ket{\Psi(0,0)}=\ket{\varphi_1}\otimes\cdots\otimes\ket{\varphi_N}$ and adiabatically changing the Hamiltonian from $H(0,0)$ to $H(\beta,t)$. In fact, the first step of the procedure -- changing the Hamiltonian from $H(0,0)$ to $H(0,t)$ -- corresponds to applying a unitary transformation 
$U=e^{itK_2}$ to $\ket{\Psi(0,0)}$. 
This transformation can be implemented \emph{exactly} in time $t$ by evolving with $K_2$ (rather than $H$). Alternatively, using the fact that $K_2$ is a sum of local commuting terms, $U$ can be decomposed into a finite-depth local unitary circuit (where the number of layers only depends on the structure of the interaction), which can be realized exactly on a digital quantum computer or simulator in constant time.

The task that needs to be implemented adiabatically is the second part of the preparation, that is, the interpolation from $\ket{\Psi(0,t)}$ to $\ket{\Psi(\beta,t)}$. Generally, the time required for a faithful adiabatic evolution will depend on the magnitude of the spectral gap of $H(\beta',t)$ along the interpolation $\beta'\in[0;\beta]$. As it turns out, for the given type of interpolation, we can devise an efficient way of checking the presence of such a gap numerically, which we present in Section~\ref{sec:section_gaps}.
Once we have established a lower bound on such a gap, we can use any standard bound for adiabatic evolutions \cite{Jansen_2007, albash_lidar}, which gives an adiabatic runtime scaling as $T = O(N^2 \Delta^{-3} \epsilon^{-1}$), with $\Delta$ a lower bound on the gap, and $\epsilon$ the error in the final state.

Moreover, for the situation at hand, we can improve the scaling of the adiabatic preparation by making use of the locality of the Hamiltonian, combined with the version of the adiabatic theorem proven in Ref.~\cite{Ge_2016}. To this end, we construct an alternative interpolation from $H(0,t)$ to $H(\beta,t)$ where in each step, we only change \emph{one} of the terms in the Hamiltonian; importantly, the method derived in Section~\ref{sec:section_gaps} to prove gaps still applies in that case. By changing the imaginary time in a suitably smooth way along this interpolation, we then obtain that the adiabatic time required per Hamiltonian term changed scales logarithmically with the desired accuracy. We can now concatenate the interpolation for all the individual Hamiltonian terms. However, we would still have a Hamiltonian acting on the whole lattice, which will translate in an overhead in the runtime if we want to devise a digital algorithm. To overcome this, we use the fact that changes performed on distant terms can be equally well carried out in parallel due to an effective light cone through Lieb-Robinson-bounds~\cite{LRbounds,Ge_2016}. This means that at each step we only work with Hamiltonians supported on a region of size poly-logarithmically in the system size. 
Combining all these results leads to a preparation scheme for $\ket{\Psi(\beta,t)}$ where the adiabatic time scales \emph{poly-logarithmically} with the desired accuracy and the problem size $N$, and thus exponentially better than other known methods for preparing MPS and PEPS. For a more technical discussion on this preparation method we refer to the reader to \cite{Ge_2016}. The bounds for the continuous-time version of the algorithm can essentially be derived from the ones presented in the original work.

% -------------------------------------------------------------
\subsection{Connection to tensor networks}\label{section_peps}

Let us now show that the states \eqref{Psibetat} can be efficiently described as PEPS \cite{cirac2020matrix} with the same connectivity as ${\cal G}({\cal V},{\cal E})$   and a bond dimension $D$ which  is upper bounded by a function of $r_1$, $r_2$, $z$, and $d$, but which does not depend on $N$ or $M$. To see this, let us consider an edge $(i,j) \in \cal E$. Since the operators $\kappa_{1,n}$ commute pairwise, we can express the product of operators that act on both $i$ and $j$ as a single one, $\prod_{i,j \in \lambda(\kappa_{1,n})} e^{i t \kappa_{1,n}} = e^{it \sum_{i,j \in \lambda(\kappa_{1,n})}\kappa_{1,n}} $. We can bound the number of terms that appear in the sum in the exponent by the number of operators that act on qudit $i$. Since the individual operators $\kappa_{1,n}$ act on a radius $r_1$, note that $e^{it \sum_{i \in \lambda(\kappa_{1,n})}\kappa_{1,n}}$ acts on, at most, all the qudits that are at a distance less or equal than $2r_1$ from qubit $i$. The number of such neighbors can be bounded  by
\begin{equation}
x \le	z \cdot \sum_{i=1}^{2r_1-1} \left(z-1\right)^{i} = z\frac{1-(z-1)^{2r_1}}{2-z} \  =  \ O(z^{2r_1})\ .
\end{equation}
Finally, note that an operator that acts on $x$ qudits increase the bond dimension at most by $O(d^{x})$.\footnote{This can be easily checked by iteratively performing a singular value decomposition on the operator and representing it as a Matrix Product Operator (MPO) acting on the $x$ qubits.} 
Iterating this for every edge (which overcounts interactions) gives an upper bound $O(d^{z^{2r_1}})$ for the required bond dimension.
Since a similar argument is valid for the operators $\kappa_{2,m}$, we conclude that the states of the form \eqref{Psibetat} can be described by a tensor network of bond dimension at most $O(d^{z^{2(r_1+r_2)}})$. Importantly, the bond dimension remains bounded independent of the system size since $r_{1}$, $r_2$ and $z$ are size-independent.

%In the following, we particularize the above results to standard PEPS -- in particular, injective MPS and PEPS, and a set of (possibly) non-injective PEPS with unique ground state. 

In the following, we particularize the above results to standard PEPS.
In particular, we will show that it contains all injective MPS and PEPS, as well as a PEPS corresponding to classical models, which (e.g.\ on the square lattice) are described by non-injective PEPS with a unique ground state. On the other hand, we generally cannot expect that $G$-injective PEPS -- that is, those which can exhibit topological order -- are of this form, regardless of boundary conditions and thus ground space degeneracy, as they are not connected adiabatically to product states; however, this can be easily remedied by choosing an RG fixed point in the corresponding phase as an initial state, rather than a product state, allowing for all findings in this paper to be carried over with only minor adaptions.

We will in the following consider regular lattices in one or higher dimensions. In the first case, we have MPS, and for the higher dimensional case, we have PEPS.

% -------------------------------------------------------------
\subsubsection{Injective MPS}

Matrix Product States (MPS) are the simplest TN \cite{Werner,cirac2020matrix}. They can be written as
 \be
 |\Psi\rangle = \sum_{s_1,\ldots, s_N=0}^{d-1} A_{s_1}^1 \ldots A_{s_N}^N |s_1,\ldots, s_N\rangle.
 \ee
We consider a special subclass, so called injective MPS. They fulfill $d=d_0^2$, and are constructed with two qudits $L_n$ and $R_n$ each of dimension $d_0$ per vertex, and
 \be
 \label{MPS}
 |\Psi\rangle = \bigotimes_{n=1}^N Q_n |\Phi\rangle \ ,
 \ee
where $|\Phi\rangle$ is a state where, at each vertex, the qudit $R_n$ is in a maximally entangled state with the qudit $L_{n+1}$ on the vertex to its right (and thus $L_n$ in a maximally entangled state with $R_{n-1}$), and $0<Q_n\le \Id$ are invertible operators that transform $\mathds{C}^{d_0}\otimes \mathds{C}^{d_0}\to \mathds{C}^{d}$. Generically, MPS become injective by blocking $\le D^4$ original qudits \cite{SanzWieland}.

\begin{figure}
    \includegraphics[width=\linewidth]{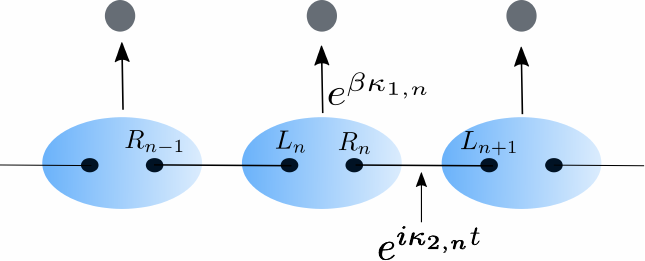}
    \caption{MPS construction in terms of the operator $e^{i \kappa_{2,n}t}$, that creates an entangled pair between sites $R_n$ and $L_{n+1}$, and $e^{\beta \kappa_{1,n}}$, that maps the virtual sites $L_n, R_n$ to the physical qudit at position $n$.s }
    \label{fig:mps}
\end{figure}

MPS obviously fall within the family of states defined in \eqref{Psibetat} for the special case of a 1D lattice as graph. Here, for each $n$, $\kappa_{2,n}$ acts on $R_n$ and $L_{n+1}$ in such a way that $e^{i\kappa_{2,n}}t$ creates the maximally entangled states on those qudits. The operator $\kappa_{1,n}$ are given by $\kappa_{1,n}=-\log(Q_n)/\beta$ and act on a single vertex each. This is illustrated in Figure~\ref{fig:mps}.

% -------------------------------------------------------------
\subsubsection{Injective PEPS}

Injective PEPS are the generalization of \eqref{MPS} to higher spatial dimensions. The state has the same form, but now there are $z_n$ qudits at site $n$, where $z_n$ is the coordination number of vertex $n$ (i.e., the number of edges connected to the vertex). The state $\ket\Phi$ contains entangled pairs along all possible vertices of the lattice. As for MPS, one can readily see that these states lie in the family defined in \eqref{Psibetat}.

% -------------------------------------------------------------
\subsubsection{PEPS corresponding to classical models}

Let us consider a classical model in a lattice. These ground states can be described by PEPS, which might be non-injective, but whose parent Hamiltonian has a unique ground state \cite{perezgarcia2007peps}. Consider a Gibbs state of some classical Hamiltonian $H_{\rm cl}(s_1, ..., s_n) = \sum_{\langle i_1 \ldots i_k \rangle } h_{i_1 \ldots i_k}(s_{i_1} \ldots s_{i_k})$, where $s_i \in \{ 0,1,\ldots,d-1 \}$, and $\langle i_1 \ldots i_k \rangle $ denotes a the regions of neighbouring particles coupled by the interaction. We define the state
\begin{equation}
   \ket{ \Psi} = \frac{1}{Z} \sum_{s_1, \ldots , s_n} e^{-\beta H_{\rm cl}(s_1, ..., s_n)/2} \ket{s_1, ..., s_n}\ ,
    \label{eq_ising_psi}
\end{equation}
where $Z$ is the classical partition function. We can rewrite this state as follows \cite{Verstraete_2006}:
\begin{equation}
    \ket{ \Psi}  = \frac{1}{Z} e^{-\beta \hat H_{\rm cl}/2} \left(\frac{1}{\sqrt{d}}\sum_{s=0}^{d-1} \ket{s}\right)^{\otimes n},
\end{equation}
where $\hat H_{\rm cl}$ is an operator with eigenstates $\ket{s_1, ..., s_n}$ and eigenvalues $H_{\rm cl}(s_1, ..., s_n)$. This state is of the type of Eq.~\eqref{Psibetat}, where $K_1 = H_{\rm cl}/2$ and $K_2 =0$. A description of these states in terms of PEPS can be easily obtained. For the special case of two-body nearest-neighbor interactions, we get a bond dimension equal to the dimension of the physical degree of freedom, see~\cite{Verstraete_2006}. 

% ================================================================
\section{Gaps and Expectation values}\label{section_GapsCorr}

In this section we establish our two key results, that find application in both efficient preparation protocols for the states \eqref{Psibetat} as well as for their verification. First, we develop an effective method to compute lower bounds to the gap of the parent Hamiltonians \eqref{eq:def-Hbetat} for some range of parameters $t,\beta$. This will ensure that the corresponding ground states can be efficiently prepared with the adiabatic algorithm presented in the next section. Second, we determine the expectation values of a complete set of operators in those states, so that one can use them to check that the state has been successfully prepared.

% -------------------------------------------------------------
\subsection{Gaps} \label{sec:section_gaps}

We will now explain how to efficiently obtain a lower bound on the gap of the parent Hamiltonians constructed in the previous subsection, for specific points in parameter space as well as uniform bounds for a whole parameter regime. To start with, note that $H(\beta=0,t)$ is a sum of commuting projectors $h_j$ for any $t$, and thus has a gap $\Delta=1$. It thus remains to supply methods to bound the gap in the case where $\beta>0$.

Let us first discuss how to obtain such a bound for a specific point $H(\beta,t)=\sum h_j$ in parameter space.
To this end, consider the semidefinite program (SDP):
\begin{subequations}\label{sdp}
\begin{equation}
\delta =  \max_{a_{ij}, c_{ij}}  x    \label{sdp1} 
\end{equation}
subject to
\begin{align}
\!\!\forall i\neq j:\hspace*{5.7cm}&\nonumber\\
h_ih_j+h_jh_i+ a_{ij}h_i^2 + a_{ji}h_j^2
  -c_{ij} h_i -c_{ji}  h_j & \succeq 0 
\label{hnm} 
\\ 
 \forall i\!: \quad \sum_{j \ne i} a_{ij}  &= 1 \label{sdp3} \\ 
 \forall i\!:  \quad \sum_{j \ne i} c_{ij}  & = x  \label{sdp4} 
\end{align}
\end{subequations}

For any feasible point of the SDP (and in particular the optimum in \eqref{sdp1}), summing \eqref{hnm} over all $i\ne j$ yields (counting each pair twice, and up to a factor $2$)
\begin{equation}
\label{eq:sdp-sum-of-all}
\sum_{i\ne j}h_ih_j + \sum_i h_i^2 - x \sum_i h_i\ \succeq 0 \, ,
\end{equation}
or equivalently
\begin{equation}
 H^2 - x H \succeq 0 \, ,
\end{equation}
which says that $H$ has no eigenvalues in the interval $(0;x)$. Since the ground state energy is $0$ by construction, this implies the existence of a spectral gap $\Delta\ge\delta$ above the unique ground state.

Since  \eqref{sdp} is an SDP with the dimension of the constraints independent of $N$,
it can be solved efficiently. Note that the SDP can be simplified considerably by setting
\begin{equation}
\label{sdp-only-overlapping}
a_{ij}=c_{ij}=0\mbox{\quad for\ }\lambda(h_i)\cap\lambda(h_j)=\emptyset
\end{equation}
and observing that \eqref{hnm} is trivially satisfied in those cases, leaving a number of equations linear in $N$. Another possible simplification amounts to first solve for each \emph{individual pair} $i\ne j$  the SDP which minimizes $a_{ij}+a_{ji}$ subject to $h_ih_j+h_jh_i + a_{ij}h_i^2 + a_{ji} h_j^2$ $\succeq 0$; then, if for all $i$, there exists some $0<y_i\leq 1$ such that $\sum_{j\ne i}a_{ij}<y_i$, then there is a gap (since $h_i$ is relatively bounded by $h_i^2$, so we can add positive contributions to both $a_{ij}$ and $c_{ij}$ while still satisfying \eqref{hnm}); a bound on the gap can either be computed directly from $y_i$ or through the SDP \eqref{eq:sdp-sum-of-all} while keeping $a_{ij}$ fixed.
Finally, note that for $\beta=0$ (where the $h_i$ are commuting projectors), condition \eqref{hnm} holds for any choice of  $a_{ij}=c_{ij}$, and thus indeed gives $\delta=1$.

Better bounds on the gap can be obtained by relaxing \eqref{hnm} to only hold when summed over specific groups of terms (which still implies \eqref{eq:sdp-sum-of-all}); a natural such case would be the variant of the SDP obtained by first grouping adjacent terms $\tilde h_i=\sum h_i$ (with the sum over terms in some neighborhood) and then setting up the SDP for the $\tilde h_i$. Furthermore, one can replace $h_i$ (also after blocking) by projectors with the same range as $h_i$, since those relatively bound $h_i$ and thus have a system-size independent gap if and only if the original Hamiltonian does.

Having described a way how to efficiently obtain a lower bound to the gap for a given point $H(\beta,t)$, how can we use this to build methods for certifying a gap over a whole range of parameters? The idea is based on the continuity of the SDP conditions (which should come as no surprise, given the finite dimension and smooth dependency of $\beta$ of all objects involved). Given a certified gap $\delta$  for some $H(\beta,t)$
using the SDP \eqref{sdp} (with corresponding optimal parameters $a^\ast_{ij}$ and $c^\ast_{ij}$),
we show in Appendix \ref{appendix_continuity_gap} that the SDP for $H(\beta+\tau,t)$ has a feasible point with $a'_{ij}=a^\ast_{ij}$, and where 
\begin{equation}
c_{ij}' =  e^{-2|\nu_j|\tau} c^\ast_{ij} - (1-e^{-2|\nu_j|\tau}) - a^\ast_{ij}\big(e^{-2|\nu_j|\tau}-
e^{-2|\nu_i{\setminus}\nu_j|\tau}\big)
\end{equation}
for $0\le a_{ij}\le 1$, and a variant thereof if $a_{ij}<0$ or $a_{ij}>1$, see Eqs.~\eqref{app-sdp-cprime}. Importantly, $c'_{ij}$ changes uniformly continuous as $\tau$ is increased starting from $\tau=0$: Thus, this
allows one to obtain a lower bound $\delta(\tau)$ on the gap of $H(\beta+\tau,t)$ by virtue of \eqref{sdp4}; importantly, the lower bound is uniform for a given interaction geometry and independent of the system size or the specific chosen model, and changes uniformly continuous with $\tau$. \iffalse \estherc{Couldn't $\tau$ depend on $N$?} \flavioc{In general yes. But for TI systems we can use the bootstrap method to argue that, if one finds a gap for a given $\tau$ and fixed $N$, then the state for the same $\tau$ is gapped for any $N$. Actually, I would propose to move the symmetrization paragraph below to up here. And use it to argue what I just said.}\fi  We now take the $\tau_0$ for which the lower bound closes, i.e. $\delta(\tau_0 ) = 0$, and re-run the SDP for $H(\beta+\tau_0,t)$: If that SDP returns a gap as well, this proves that $H(\beta+\tau,t)$ is gapped for all $0\le\tau\le\tau_0$.
By starting from $\beta=0$ (for which \eqref{sdp} trivially holds), we can thus establish the existence of a gap (and obtain explicit lower bounds to it) for a finite regime $0\le\beta\le\beta_0$ by evaluating the SDP \eqref{sdp} only at a finite number of points (where the bound can be improved by increasing the density of points). Note that for translational invariant systems, we can show that the SDP from \eqref{sdp} can be ``symmetrized" (see Appendix \ref{appendix_symmetrization}). This means that for a given feasible point, with values of the variables $(a^*_{ij}, c^*_{ij})$, we can find another feasible point by averaging $a^*_{ij}$, $c^*_{ij}$, such that all coefficients $a_{ij}$ and $c_{ij}$ are equal for all pairs of terms in \eqref{sdp1}. Moreover, this solution provides the same value of $\delta$. This directly shows that in this case $\delta$ does not change with the system size, thus recovering the result first showed in Ref. \cite{Werner} about the existence of a gap independent of the system size. Therefore, for TI systems the size of the neighbourhood $\tau$ is independent of the system size.

Let us now briefly discuss the suitability of the method for the TN of the previous section. For translational invariant (or suitably uniform) injective MPS, it was proven by Fannes, Nachtergaele, and Werner~\cite{Werner} that the parent Hamiltonian is always gapped. Indeed, they showed that by blocking a number of sites proportional to the correlation length (for a fixed bond dimension), which in turn can be bounded as a function of the gap, and replacing the blocked Hamiltonian $\tilde h_i$ by projectors, Eq.~\eqref{hnm} is fulfilled (with all $a_{ij}$ and $c_{ij}$ equal, as mentioned in the previous paragraph) -- and thus the SDP will yield a gap $\delta>0$ -- already with the restriction \eqref{sdp-only-overlapping} imposed.
In higher dimensions, an analogous result has been obtained for PEPS which are unique ground states of local Hamiltonian (in particular, injective PEPS)~\cite{Kastoryano_2018}: Whenever the Hamiltonian is gapped in the thermodynamic limit, the SDP with condition \eqref{eq:sdp-sum-of-all} will be satisfied by the projector-valued Hamiltonian obtained after blocking a number of sites which only depends on the gap and the geometry of the system, but not on the system size or details of the model.

Finally, regarding the PEPS corresponding to classical models in dimensions higher than one, for sufficiently high temperatures (small $\beta$) the SDP method will give a gap (as the Hamiltonian at $\beta=\infty$ is trivial). Note that the corresponding classical model may have a phase transition at sufficiently low temperatures, which implies that the correlation length will diverges and thus the gap will vanish in the thermodynamic limit ($N\to\infty$). Thus, the SDP automatically also allows to determine an upper bound to that critical temperature, which will yet again get more accurate as we consider larger regions, both by blocking and by relaxing (or omitting) the restriction~\eqref{eq:sdp-sum-of-all}. Thus, the continuity bound of Appendix~\ref{appendix_continuity_gap} at the same time provides a means of determining upper bounds on the critical temperature of classical statistical models.

% -------------------------------------------------------------
\subsection{Expectation values} \label{sec:section_expectation}

Computing expectation values of ground states of local Hamiltonians is hard in general \cite{localQMA}. However, for ground states of frustration-free Hamiltonians, there are certain observables for which it is straightforward to compute such values. In this section, we find a complete set of operators for which this can be done and which will be on the basis of the verification protocols presented below.

As argued in Section \ref{section_parent}, for $\ket\Psi$ given in \eqref{Psibetat} we have $h_n|\Psi\rangle=0$, and thus trivially $\langle \Psi|h_n|\Psi\rangle=0$. We will now define a set of observables for which one can also compute the expectation value in $\ket\Psi$. We will restrict to qubits ($d=2$) and will take $|\varphi_j\rangle=|0\rangle$, although it is straightforward to extend the method to qudits and other states. We will denote 
the Pauli operators acting on the $j$-th qubit, 
$j=1,\ldots,N$, 
by $\sigma_{\alpha}^j$ with $\alpha=x,y,z$;
 for instance, $\sigma_z|0\rangle=|0\rangle$. 

The key idea is to notice that for any $\lambda\subset {\cal V}$,
 \be
 \label{Olambdaqprop}
  O_\lambda |\Psi\rangle = \big(\bigotimes_{j\in\lambda} |0\rangle_j\langle 0| \big) O_\lambda |\Psi\rangle \ ,
 \ee
where
 \be \label{eq:o-lambda}
 O_\lambda = \prod_{n\in \mu(\lambda)} e^{-i \kappa_{2,n} t} \prod_{m\in \nu(\lambda)} e^{-\beta \kappa_{1,m}}
 \ee
with $\mu(\lambda)=\bigcup_{j\in\lambda}\mu_j$,
$\nu(\lambda)=\bigcup_{j\in\lambda}\nu_j$,
and where the sets $\mu_j,\nu_j$ have been defined in Section \ref{section_parent}.

The first set of operators is defined as $Z^+_\lambda = (Z_\lambda + Z_\lambda^\dagger)/2$  and 
$Z^-_\lambda = (Z_\lambda - Z_\lambda^\dagger)/2i$, where 
 \be
 \label{Zlambda}
 Z_\lambda = O_\lambda^{-1} \left(\bigotimes_{j\in\lambda}\sigma_z^j \right) O_\lambda
 \ee
 -- that is, $\ket{\Psi}$ is a right (left) eigenvector of $Z_\lambda$ ($Z_\lambda^\dagger$), using \eqref{Olambdaqprop}.
 We then have
 \bse
 \bea
 \label{Zplus}
 \langle \Psi| Z^+_\lambda |\Psi\rangle &=& 1\ ,\\
 \langle \Psi| Z^-_\lambda |\Psi\rangle &=& 0\ ,
 \eea
 \ese
and thus $Z^\pm_\lambda$ have fixed expectation values.

The second set is more ample. Given any $P$ supported in $\lambda\subset {\cal V}$ with the property that there exists $j \in \lambda $ such that $\bra{0}  P  \ket{0}_j = 0$, we define
 \be
 \label{eq:q-lambda}
 Q_\lambda = O_\lambda^\dagger P O_\lambda.
 \ee
Again, using \eqref{Olambdaqprop} we can compute the expectation value of those operators
 \be
 \langle \Psi| Q_\lambda |\Psi\rangle = 0.
 \ee
 
The set of operators defined above, taken jointly for all $\lambda\in\mathcal V$, is complete in ${\cal A}$ in the sense that their expectation values completely determine the state. To show that, we just have to devise a subset of $4^N$ linearly independent operators from that set.

To this end, let us start from the set of all operators $P$ which are a product of Pauli and identity operators, and associate to each of them an operator $Q\equiv Q(P)$ using one of the constructions above.
Each of these operators can be written as 
 \be
 \label{eq:P-pauli-substring}
 P=\bigotimes_{j\in \lambda} \sigma^j_{\alpha_j}\ ,
 \ee
where $\alpha_j=x,y,z$, and  $\lambda\subseteq {\cal V}$ is the set of sites on which $P$ acts non-trivially.
In case $\alpha_j=z$ for all $j\in\lambda$, we define $Q(P)=Z^+_\lambda$; otherwise, $Q(P)= Q_\lambda$, Eq.~\eqref{eq:q-lambda}. Finally, for $P=\Id$, let $Q(P)=\Id$. 
In this way, starting from all products of Pauli operators and the identity, we have obtained a set of operators $Q_n$, 
$n=1,\ldots,4^N$. 
This set is linearly independent iff the matrix $B_{n,m}={\rm tr}(Q_n^\dagger
Q_m)/2^N$ is not singular. Trivially, for $\beta=0$, $B_{n,m}=\delta_{n,m}$ and thus not singular. Since the operators $O_\lambda$ used to define the map $P\mapsto Q$ are analytic in $\beta>0$, the determinant of $B_{n,m}$ will be analytic as well, and thus it can only vanish at countably many points, all of which are isolated. Thus, generically it will be non-zero and, in the possible measure-zero cases where it is can be circumvented by taking a closeby value of $\beta$.

The fact that the set of operators $Z_\lambda$ and $Q_\lambda$ is (over)complete means that any observable can be expanded as a linear combination thereof. If we are able to obtain the corresponding coefficients, then we will be able to compute the expectation value of all physical quantities. In practice, this will be difficult since, due to the fact that the operators $O_\lambda$ non-trivially overlap with each other, we will typically need an exponential number of terms in the expansion. Nonetheless, there may be a way of truncating that expansion, which would give lead to new algorithms in terms of tensor networks. Furthermore, since we know the expectation values of a basis of operators, we possess full tomographic information on state. However, as before, it is not useful to compute other expectation values. In Appendix \ref{appendix_norms_Olambda} we show that the norms of the observables $Q_\lambda$ decay at most exponentially with their support size, and thus they are guaranteed to have bounded (polynomially-decaying) norm when the size of the support is fixed (at most $\mathcal{O}(\log N)$).

The operators $Q_\lambda$ are supported on a set of vertices that is larger than $\lambda$ (roughly speaking, on all vertices that lie at a distance up to $2r$ from that set). It is possible to define other observables which have a smaller support and for which we can still compute the expectation values. This is relevant for more practical applications, like the verification protocols introduced below, where we want to make statements about measurements performed within the support of $Q_\lambda$, and we want them to include as few qubits as possible. For that, given $j\in{\cal V}$ and an operator $P'$ supported on $\lambda$ with $j\notin\lambda$, we define
 \bse
 \label{Qjx}
 \bea
 Q_{j,1}&=&O_j^\dagger \sigma_x^j P' O_j\ ,\\
 Q_{j,2}&=&O_j^\dagger \sigma_y^j P' O_j\ ,\\
 Q_{j,3}&=&O_j^\dagger (\Id-\sigma_z^j) P' O_j\ ,
 \eea
 \ese
 which only act on the joint support of $O_j$ and $P'$.
Again, using \eqref{Olambdaqprop} we find the expectation value of those operators to be 
 \be
 \langle \Psi| Q_{j,\alpha} |\Psi\rangle = 0.
 \ee
 
In particular, if we choose $P'$ to be an arbitrary Pauli product, then $Q_{j,1}$ and $Q_{j,2}$ can be used to replace the operators $Q_\lambda$ from Eq.~\eqref{eq:q-lambda} in our complete set of operators. At the same time,  the operators $Q_{j,1}$ and $Q_{j,2}$ are still products of Pauli operators almost everywhere, except on the support of $O_j$ which has a fixed size. This means that each $O_{j,1}$, $O_{j,2}$ can be estimated efficiently (i.e.\ with a number of (Pauli) measurements which only depends on the accuracy but not on the system size), and the only remaining operators in the complete set which cannot be estimated efficiently individually are the $Z_\lambda^\pm$.

In summary, we have defined a set of observables whose expectation values are either zero or one. We can choose a subset thereof where $|\lambda|\le c$, with $c$ a constant independent of $N$. In such a case, since we know $O_\lambda$ we can efficiently write those observables as linear combinations of Pauli operators in the support of $O_\lambda$ (like \eqref{Zlambda}), or even smaller (like \eqref{Qjx}). Note that the norms of these observables will also be efficient to compute in these cases, as we show in Appendix \ref{appendix_norms_Olambda}. 

% ================================================================

\section{Verification schemes} \label{section_verification}
In this section we discuss different ways of exploiting the state preparation procedure for the state \eqref{Psibetat} as a verification scheme.
First, we will analyse a quantum state verification method and show how to certify that the state has been created successfully by performing local measurements and using the fact that there exists a parent Hamiltonian that is both gapped and frustration free \cite{Cramer_2010}. 
Then, we will consider the scenario of classical verification of quantum computation, where the goal is to make sure that someone else is in possession of a quantum computer solely through classical communication~\cite{reichardt2012classical, reichard2013nature, aaronson2016complexitytheoretic, Boixo_2018,mahadev2018classical, brakerski2019cryptographic}. We will restrict here to the case of qubits, although the arguments can be easily extended to qudits. 

% -------------------------------------------------------
\subsection{Quantum state verification} \label{section_quantum_verification}

Unique ground states of frustration free Hamiltonians, like the ones we are dealing with here, can be trivially certified with local measurements. This is achieved by just performing local quantum tomographies to make sure that the expectation values $\langle h_j\rangle=0$ for all $h_n$ defined in~\eqref{hj}. Indeed, if this is the case, then $\langle H\rangle=0$ [cf.~\eqref{eq:def-Hbetat}], and since $H\ge 0$ and has a unique ground state, this implies that the measurement must have been performed on that state.

In practice, since we can only perform a finite number of measurements, the estimates for $\langle h_j\rangle$ will not be exactly zero; additionally, measurement errors will give rise to errors in those quantities as well. However, one can still estimate the success probability of the preparation in different ways. The most straightforward one is to relate the obtained expectation value of $\langle H\rangle=\mathrm{tr}(H\rho)$ (with the corresponding estimated error) to the overlap of the state we have prepared, $\rho$, and the target ground state $\ket\Psi$. It is straightforward to show that the fidelity obeys
 \be
 \label{bound}
 \langle \Psi|\rho|\Psi\rangle \ge 1- \frac{\tr (H\rho)}{\Delta} \ge 1- \frac{\tr (H\rho)}{\delta}\ ,
 \ee
where $\Delta$ is the spectral gap of $H$, and $\delta$ the lower bound obtained in the previous section. Thus, if we can obtain this expectation value (with an error bar) and we compute $\delta$, we directly obtain a lower bound to the overlap.

Neglecting measurement errors, for a finite number of measurements, the estimate of $ \langle \Psi|\rho|\Psi\rangle $ will have some error bar.  
To use the bound \eqref{bound}, we need that the error in $\mathrm{tr}(H\rho)=\langle H\rangle=\sum \langle h_j\rangle$ is sufficiently below $\delta$. Assuming that we perform independent measurements and thus that we have independent errors with same standard deviation $\sigma$ for the estimator of all the $h_j$, we have that 
the error $\epsilon_j$ for each term must satisfy $\delta/\sqrt{N}\sim\epsilon_j \sim \sigma /\sqrt{L_j}$,  where $L_j$ is the number of measurements performed to estimate $\langle h_j\rangle$. If, in addition, we assume $L_j = L_{j'}$, we obtain a conservative estimate for the total number of measurements of
$L_\mathrm{tot} \lesssim N L_j \sim \sigma^2N^2/\delta^2$. Note that instead of performing independent measurements, one could measure qubits belonging to non-overlapping regions in parallel \cite{Cotler_2020}, which might lead to a reduction of the total number of rounds.

The verification can also be analyzed as an adversarial game,
where the 
prover prepares a state, $\rho$, gives it to verifier, and claims that it is indeed $\Psi$. The verifier can then perform measurements to gain confidence that this is indeed the case. Such a protocol as been analyzed in Ref.~\cite{Hangleiter_2017} by assuming that the verifier can measure by projecting in the basis of eigenstates of the local operators of the Hamiltonian $h_j$. In Appendix \ref{appendix_bounds_verification}, we perform a similar analysis, but assuming that the verifier can only perform Pauli measurements, which might be a more realistic assumption for current experimental setups.

% -------------------------------------------------------
\subsection{Classical verification of quantum computation}

Let us now turn towards a different kind of verification, in which the verifier is fully classical and communicates with the quantum prover through a classical channel~\cite{reichardt2012classical, reichard2013nature,mahadev2018classical, brakerski2019cryptographic,  aaronson2016complexitytheoretic, Boixo_2018}. Such protocols can also be formulated as an adversary game: Here, the prover claims that he can efficiently carry out quantum computations on a quantum computer. The verifier has to make sure that this is the case by communicating classically with the prover, that is, asking him to perform certain tasks on his quantum computer and report the results. Of particular interest is the case where both prover and verifier have limited additional resources, namely 
they can perform classical computations with a computational time that scales at most polynomially with the number of qubits. In this setting, the verifier can pose challenges to the prover which he can only accomplish if he has a quantum computer, but not with his limited classical resources, and the challenge is to find a way which allows the verifier with her limited classical resources to verify this. 

Recently, several protocols achieving this task have been proposed whose security is based on standard complexity assumptions \cite{mahadev2018classical, brakerski2019cryptographic, aaronson2016complexitytheoretic}. While the first protocol \cite{mahadev2018classical, brakerski2019cryptographic} is most adequate for fault-tolerant quantum computers, the latter \cite{aaronson2016complexitytheoretic} is very attractive since it can already be used to verify existing NISQ devices \cite{arute2019quantum}. However, it requires the verifier to carry out classical computations whose runtime scales exponentially with the number of qubits, though with a relatively small exponent which makes it comparatively practical. Apart from their use to certify quantum computers that can be only used remotely by classical means, one of the most appealing applications of such protocols is in the context of certified random number generators \cite{brakerski2019cryptographic}.

Ground states of frustration-free quantum Hamiltonians that can be efficiently prepared, like the ones presented in this work, may offer an alternative way for this kind of verification; specifically, one can exploit the task of reproducing correctly the expectation values of the observables introduced in Section \ref{sec:section_expectation} as a challenge for the prover. In the following, we will describe such verification protocols, and analyze their security and the underlying complexity theoretic assumptions in different regimes. As we will see, the straightforward application of this idea requires exponential time. More efficient versions of the protocol are possible, but the underlying complexity assumptions are less tangible and thus, their security remains unclear.

% -----------------------------------------------------
\subsubsection{Protocol}

The verification protocol consists of three steps: 
(\emph{i})~The verifier sends the prover instructions for preparing the state $\ket{\Psi}$.
 (\emph{ii})~This step consists of $R$ rounds: in each round, the verifier sends the prover a set of observables; the prover then prepares the state $\ket\Psi$, measures the observables, and reports the outcome. 
 (\emph{iii}) The verifier
performs certain checks on the accumulated measurement outcomes to verify that the prover has indeed prepared the state $\ket\Psi$ and is thus in possession of a quantum computer.

For step (\emph{i}),  the verifier just has to give the circuit that prepares the state to the prover, which she can e.g.\ obtain by trotterizing the adiabatic evolution. Alternatively, she can directly give the time dependent Hamiltonian $H(t,\beta)$, together with the initial states $\ket{\varphi_i}$, to the prover, e.g.\ in case he is in possession of an 
analog quantum computer.  An honest prover  will then be able to efficiently prepare the state $\ket\Psi$ in a time that scales as $\log(N)$, as discussed in Section~\ref{section_Adiabatic}.

For each round of step (\emph{ii}), the verifier sends the prover a list of bases $\bm\alpha=(\alpha_1,\dots,\alpha_N)$, $\alpha_j=x,y,z$, in which the individual qubits should be measured; the $\bm\alpha$ will be generally drawn at random from some distribution which is dictated by the verification step (\emph{iii}). The prover then prepares the state and measures qubit $j$ in the Pauli basis $\alpha_j$, and obtains results $\bm s = (s_1,\dots,s_N)$, $s_j=\pm1$. For an honest prover, these results are drawn from a distribution
\begin{equation}
\label{eq:P0-distribution}
 P_0(\bm s | \bm \alpha)=  \langle \Psi| \bigotimes_{j=1}^N \frac12(\openone+s_j\sigma_{\alpha_j}^{j})
 \ket\Psi\ .
 \end{equation}
After receiving the measurement basis, the prover prepares $\ket\Psi$ and performs the measurements. Importantly, since $\ket\Psi$ can be prepared in time $O(\log(N))$, each round can in principle be carried out in time $\log(N)$ as long as the prover has the ability to measure and communicate the results in parallel. 

Step (\emph{ii}) allows for several natural generalizations. In particular, we can allow for measurements beyond the Pauli basis, we can enable the verifier to make \emph{adaptive} queries, where the choice of $\bm s$ in any round can depend on the results of the previous round, and we can split each round into several sub-rounds of communication, where the state is prepared once and then a sequence of adaptive measurements is performed on it, measuring only a subset of the qubits at each time.

In step (\emph{iii}), the verifier uses the samples obtained from the prover to compute certain quantities which serve to verify that the prover is indeed sampling from the correct distribution $P_0$. To this end, the verifier can e.g.\ use some of the quantities $Q$ constructed in Section~\ref{sec:section_expectation} for which $\langle Q\rangle=0,1$ (or variants thereof), or the Hamiltonian terms $h_j$ for which $\bra\Psi h_j\ket\Psi=0$ (which can be used to replace the $Z^-_{\{j\}}$). Let us now consider one such observable $Q$ supported on $\lambda\subset \mathcal V$. It can be decomposed as 
\begin{equation}
\label{eq:O-Pauli-expansion}
 Q = \sum_{\bm\gamma} o(\bm\gamma) \bigotimes_{j\in\lambda} 
    \sigma_{\gamma_j}^j\ ,
\end{equation}
where $\bm\gamma=(\gamma_1,\dots,\gamma_{|\lambda|})$, $\gamma_j=x,y,z$ and $o(\bm\gamma)$ are the expansion coefficients. 
The verifier can then estimate $\langle Q\rangle$ using an estimator
 \be
 \label{eq:Obar}
 \bar Q = \sum_{\bm\gamma} o(\bm\gamma) \bar s(\bm\gamma)\ ,
 \ee
where $\bar s(\bm\gamma)$ is the average outcome where the prover measured spin $j$ in the basis $\sigma_{\gamma_j}$ for all $j\in \lambda$, and an arbitrary basis for all $j\notin\lambda$;
that is, if we denote the set of rounds where $\alpha_j=\gamma_j$ for all $j\in \lambda$ by $\mathcal R(\bm\gamma)$, and 
the result of the $r$'th round by $\bm s^r=(s_1^r,\dots,s_N^r)$, we have 
 \be
 \label{eq:O-sbar}
 \bar s(\bm\gamma) =  \frac{1}{|\mathcal R(\bm\gamma)|} \sum_{r\in\mathcal R(\bm\gamma)} \prod_{j\in\lambda} s_{j}^r\ .
 \ee
If the samples have been taken according to $P_0$, 
$\bar O \to \bra\Psi O\ket\Psi$
in the limit $|\mathcal R(\bm\gamma)|\to\infty$, which can be used as a means of verification (see Appendix~\ref{appendix_bounds_verification} for a quantitative analysis). Note that alternatively, we can determine the average also using only rounds $\mathcal R(\bm\gamma)$ where sites $j \notin \lambda$ have only been measured in some specific bases.

%%%%%%%%%%%%%%%%%%%%%%%%%%%%%%%%%

As a simple example of this verification procedure, consider
a resource state for measurement-based quantum computation (MBQC), such as the cluster state~\cite{raussendorf2000quantum}, or generalizations which allow for measurement-based computation using only Pauli measurements~\cite{Miller_2016}. For those states, the preparation step is particularly easy, as they can be prepared by a single layer of commuting unitaries (i.e.\ $\beta=0$) from a product state. They are frustruation free ground states of local Hamiltonians $H=\sum h_i\ge0$, $h_i\ket{\Psi}=0$, a property which allows for easy verification from the measurement statistics. And finally, we know that adaptive measurements in a suitable basis (Paulis and $\pi/4$ rotated states in the xy plane for the cluster state, or only Pauli measurements for the aforementioned generalization) are universal for quantum computation, making it impossible for a prover not in possession of a quantum computer to produce the correct distribution in such an interactive protocol, 
and at the same time imposing limitations on potential cheating strategies, as we will discuss in the following.

% -----------------------------------------------------
\subsubsection{Complexity Analysis}

Let us now analyze under which conditions the protocol allows the verifier to conclude that the 
prover must indeed be in possession of a quantum computer, what potential classical cheating strategies might be, and what limitations to such strategies exist.

\paragraph{Sampling from the quantum distribution is hard.\label{para:hardness-true}}
The first cheating strategy would be to find a way to classically sample from the correct distribution $P_0$, Eq.~\eqref{eq:P0-distribution} -- in that case, the verifier would have no way of detecting this. However, there are several strong complexity-theoretic arguments against that. First, note that 
-- as already mentioned above --
the adaptive version of the protocol contains measurement-based quantum computing: The cluster state is clearly in the given class (with $\beta=0$), and an adaptive protocol with $\mathrm{poly}(N)$ queries per round would implement a general quantum computation. Thus, sampling from the correct $P_0$ is impossible unless $\textsf{BQP}=\textsf{BPP}$. However, it is also known that sampling from the output of a circuit of commuting gates with random product states as input and measuring in a fixed basis (a case which is contained in our protocol for $\beta=0$) is computationally hard, as shown in \cite{Bermejo_Vega_2018}. In order to prove the hardness of this setup, in \cite{Bermejo_Vega_2018} they relate the complexity of sampling from such a circuit with the complexity of sampling from \textsf{IQP} circuits, a task that is known to be hard unless the polynomial hierarchy would collapse to its third level~\cite{Bremner_2010,Bremner_2016}.
We thus conclude that there is complexity-theoretic evidence that it is impossible for the prover to classically sample from the correct distribution $P_0$ (up to constant error) in polynomial time.

\paragraph{Reproducing all $\langle Q\rangle$ is hard.}
The second cheating strategy would be to sample from some different distribution $P'$ which is chosen such that all estimators for $\langle Q\rangle$ computed from $P'$ are correct.
Any estimator $\bar Q$, Eq.~\eqref{eq:Obar}, supported on a subset $\lambda$ of sites  can be computed in many different ways, which differ by the measurement settings over which we average for the sites not contained in $\lambda$, see Eq.~\eqref{eq:O-sbar} and the discussion below it. We will thus additionally impose that the estimator $\bar Q$ converges to the same value for all those ways to compute it; this can be easily ensured by the verifier by computing $\bar Q$ in all different ways, or just in a randomly chosen one. (Alternatively, this can be ensured by computing the marginal probability distributions in different ways and checking their consistency, that is, by checking that the measurement setting $\alpha_i$ of qubit $i$ does not affect the distribution of the measurement results $s_j$ for any of the other qubits $j\ne i$; this can be seen as a kind of non-signalling condition on the distribution.)

These conditions however imply that $P'=P_0$, the correct distribution derived from the quantum state. The reason is that for any given $\bm\alpha$, we can reconstruct $P'(\,\cdot\,|\bm \alpha)$ from the expectation values of all Pauli measurements given by arbitrary \emph{substrings} of $\bm\alpha$ (that is, where the Paulis in some positions have been replaced by identities, as in Eq.~\eqref{eq:P-pauli-substring}) through an $N$-fold Hadamard transform (using the consistency condition above); the latter, in turn, can be reconstructed from all $\langle Q\rangle$ as they are related by an invertible transformation, as discussed in Section~\ref{sec:section_expectation}.

We thus find that this is not a viable cheating strategy, since no other distribution which yields all the correct $\bar Q$ exists. Note that together with the hardness results mentioned under point \ref{para:hardness-true}, this implies that even sampling from a $P'$ which approximates $\langle Q\rangle$ for all $Q$ sufficiently well is hard. The required accuracy in $Q$ has to be chosen such that the expectation values of Pauli strings have exponential accuracy in $N$; since the number of samples required to this end is in fact determined by the latter (the transformation \eqref{eq:Obar} is exact), this generally requires an exponential number of samples.

Let us now discuss two different cases for the verification protocol and their security. First, we consider the case where both prover and verifier have bounded (polynomial) storage space, but the verification procedure can take an exponential number of rounds, and the verifier can use exponential time. In this case, the arguments from before regarding the hardness of sampling from $P_0$ or reproducing the $\langle Q \rangle$ imply the security of the protocol. Second, we consider the case in which the verifier only tests local observables, i.e., those supported on a region of at most $O(\log N)$. Note that the verification is efficient in this case, i.e., it can be carried out in polynomial time for a prover and verifier having both polynomial storage space as before. In this case, we also argue how successful cheating strategies will be likely to fail. Most interesting, if such a cheating strategy succeeded, it would imply the existence of classical algorithms for computing local observables for generic tensor network states. 

\paragraph{A secure space-bounded and exponential-time verification protocol.} In each round, the prover is only granted $\mathrm{poly}(N)$ time (or logarithmic time with suitable parallel processing power). By performing an exponential number of queries, the verifier can get an exponentially precise estimate $\bar Q$ for any of the observables $Q$ (either a randomly selected one,  or all of them); importantly, the expansion coefficient $o(\bm\gamma)$ of $Q$ in the Pauli basis, Eq.~\eqref{eq:O-Pauli-expansion}, is a trace of a product of local operators and can thus be computed in \textsf{PSPACE}, and the summands in \eqref{eq:Obar} can be sampled and added sequentially one by one; the verifier thus only requires polynomial storage space. The prover, one the other hand, cannot classically sample from the correct distribution in the available polynomial (or logarithmic) time, due to the hardness results discussed in point \ref{para:hardness-true}; at the same time, his limited memory prevents him from pre-computing all possible outcomes after having learned the adiabatic circuit for preparing $\ket\Psi$ in step (\emph{i}) (even though the exponential time used for the overall protocol would allow for it).

Let us note that what makes our setup special is not the fact that the knowledge of all $\langle Q\rangle$ allows to reconstruct the probability distribution -- such 
a complete tomography is possible in any scenario. 
Rather, 
what makes our construction special is that the verifier knows the required expectation values for all operators $\langle Q\rangle$ right away, whereas in a general tomography scheme, a computationally costly reconstruction procedure is required to obtain $P$ from measured expectation values.
However, in order to compute a general expectation value $\langle Q\rangle$, the verifier still needs to collect an exponential amount of data from the prover: While it is sufficient to sample a small number of randomly chosen $Q$'s, most $Q$'s still have an exponential number of terms in their Pauli expansion \eqref{eq:O-Pauli-expansion}.
The situation here can thus be somehow regarded as the reverse from that introduced in Ref.~\cite{aaronson2016complexitytheoretic}: While in that case, the verifier requires a polynomial number of measurements from  the prover, it takes her an exponential time to check if they correspond to the correct probability (by computing the cross-entropy). In our case, the correct expectation values can be trivially computed, but one requires an exponential number of samples to obtain them. Note that computing the coefficients $o(\bm\gamma)$ of $Q$, which are needed to estimate its expectation value from the samples, might require exponential resources as well.

\paragraph{Proposal for an efficient verification protocol and its security implications.} 
As we have seen, checking some $\langle Q\rangle$, chosen at random from the set of \emph{all} $Q$, allows to verify that the prover is in possession of a quantum computer. This, however, requires exponential resources, since a typical $Q$ involves a significant fraction of the Pauli terms acting on its support, that is, an exponential number, and moreover we require an exponential accuracy. A practical verification protocol must thus resort to testing only \emph{local} $Q$'s, that is, those which are supported on at most order $\log(N)$ sites, to an accuracy $1/\mathrm{poly}(N)$, as those can be sampled in $\mathrm{poly}(N)$ time.

Thus, the question which arises is whether there is a way for the prover to efficiently sample from a distribution $P'$ which reproduces those local $\langle Q \rangle$ correctly, up to polynomial accuracy.

First, let us notice here that, if the prover indeed measures in the Pauli basis, reaching a polynomial accuracy on local $\langle Q \rangle$ expectation values uniquely identifies the ground state $\ket\Psi$ among all multi-qubit states. The reason is that we can include the terms of the parent Hamiltonian $H=\sum h_j$ in the set of $Q$'s, and since $\ket\Psi$ is the unique ground state of $H$ and $H$ is gapped, this implies $\|\ket\Phi-\ket\Psi\|<1/\mathrm{poly}(N)$ if the $\langle h_j\rangle$ are $1/\mathrm{poly}(N)$-close, see Eq.~\eqref{bound}. Hence, to entirely establish the security of the protocol, it would be enough to prove that any distribution $P'$ reproducing $\langle Q \rangle$ with the desired accuracy must be obtainable by performing Pauli measurements on a multi-qubit state. Notice that a possible way to achieve that would be to prove a robust self-testing statement based on the $\langle Q \rangle$ expectation values (see \cite{Supic2020selftestingof} for a review of self-testing techniques).

Second, even if such a distribution $P'$ exists, there are constraints on the design of algorithms to sample from it efficiently.  For instance, one might assume that it should be possible to characterize the space of solutions of the equations $|\bar Q-\langle Q\rangle|\le 1/\mathrm{poly}(N)$, which locally constrain $P'$, use them to find ways to sample locally correctly, and patch those ways of sampling together to obtain a sampling algorithm which works globally. However, such a strategy, if based only on the final conditions on $\langle Q\rangle$, is bound to fail, unless further properties of $P_0$ are taken into account. Specifically, we can choose the $Q$ to enforce 3-SAT clauses or some other classical \textsf{NP}-hard problem (such as spin glasses on a 2D lattice), in which case such an algorithm is bound to fail as it would have to solve the \textsf{NP} problem. (It is a fingerprint of hard instances of \textsf{NP} problems that the local constraints cannot be patched together easily.) Note that also knowing the precise local reduced density matrices does not necessarily make the problem easier, as the general quantum marginal problem is \textsf{QMA}-hard~\cite{liu:marginal-qma,broadbent:marginal-qma}. 

Thus, a successful cheating strategy most likely will have to use the full knowledge of the adiabatic path used to prepare the state, or equivalently knowledge of $K_1$ and $K_2$. Note that Osborne showed that the computation of local expectation values on states that undergo an adiabatic evolution under a local gapped Hamiltonian can be achieved classically \cite{osborne2007simulating}. However, this method does not help the prover either, since it does not provide samples from the full probability distribution, which is what the prover is asked to return. Another approach could be to classically simulate the adiabatic evolution -- for instance, one could try to adapt the Monte Carlo algorithm by Bravyi and Terhal~\cite{bravyi:stoq-ffree} for the simulation of adiabatic quantum computation along a path which is both frustration free (which we have) and sign-problem free (which we don't necessarily have), followed by a measurement in the $\sigma_z$ basis (which we don't have). Indeed, it is plausible that such an algorithm will allow to correctly sample in cases where the final Hamiltonian is classical and only needs to be sampled in the $\sigma_z$ basis. On the other hand, it will likely break down if one of the conditions is not met, which will manifest itself in a sign problem in the Monte Carlo method. In fact, we cannot expect a cheating strategy based on simulating the adiabatic evolution to work, since such an algorithm (if working as desired) will precisely sample from the \emph{correct} distribution $P_0$, which we have previously established to be computationally hard. Thus, in order to attack the protocol with such a strategy, one would have to devise a method to simulate the adiabatic evolution in a way where local expectation values are reproduced correctly, yet global properties would not 
(with the goal of circumventing hardness results);
it is unclear how a route to accomplish this would look like.

\begin{acknowledgements}
We thank Adam Bouland, Zeph Landau, and Umesh Vazirani for helpful discussions. This work has received support from the European Union's Horizon 2020 program through the ERC-AdG QUENOCOBA (No.~742102) and the ERC-CoG SEQUAM (No.~863476), from the DFG (German Research Foundation) under Germany's Excellence Strategy (EXC2111-390814868), and from the Austrian Science Fund (FWF) through Project number 414325145 within SFB F7104. 
	J.T.\ thanks the Alexander von Humboldt foundation and the Google Research Scholar Program  for support.
\end{acknowledgements}

	\bibliographystyle{h-physrev}
	\bibliography{biblio}

\newpage
\onecolumngrid
\appendix

\section{Examples of set of operators $\mathbb{K}_{r,M}$} \label{appendix_operators}

In this section we give some examples on how to construct operators $K \in \mathbb{K}_{r,M}$ defined in \eqref{decomposition}. Recall that $K$ is defined as a sum of local operators $\kappa_n \in \mathcal{A}_r$, for $n=1,...,M$, such that the operators $\kappa_n$ commute pairwise.

Let us take a set of pairwise commuting unitary operators acting on at most $r_U$ sites: $\left\{ U_{\alpha}:  \quad | \lambda(U_{\alpha})| \leq r_U, \quad \left[ U_\alpha, U_\beta \right]=0 \right\}$. Note that a straightforward way of constructing commuting, Hermitian operators is as follows
\begin{equation}
    \kappa_n = \left( \prod_{\alpha: n \in \lambda(U_{\alpha})} U_{\alpha_i} \right) O_{n} \left( \prod_{\alpha: n \in \lambda(U_{\alpha})} U_{\alpha_i} \right)^\dagger,
\end{equation}
where $O_n$ is a one-body operator acting on site $n$ such that $\Vert O_n \Vert_{\infty} \leq 1$ (note that we can also take $O_n$ to act on disjoint subset of vertices $\lambda({O_n}) \subset{\mathcal{V}}$, such that $\lambda({O_n}) \cap \lambda({O_n'}) = \emptyset$). Note that the product runs over all unitary operators whose support intersect site $n$. In this case, the final operators $\kappa_n$ are labeled by a physical site, though in the general definition \eqref{decomposition} this is not necessarily the case.  Note that the conjugation by unitaries preserves the commutativity, i.e. $\left[\kappa_n, \kappa_m \right] = 0$ for all $n \neq m$, as well as the spectrum of the operators, so $\Vert \kappa_n \Vert_{\infty} \leq 1$. We can also upper bound the support size of $\kappa_n$, $|\lambda(\kappa_n)| \leq z \cdot r_U -1$.

\begin{figure}
	\begin{subfigure}{0.48\textwidth}
		\vspace{0.6cm}
		
		\includegraphics[width=0.58\linewidth]{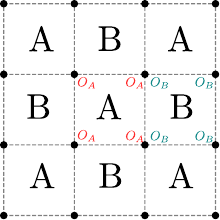}
		\caption{Toric-code unitary operators.} \label{fig:a_toric}
	\end{subfigure}%
	\hspace*{\fill}   % maximize separation between the subfigures
	\begin{subfigure}{0.48\textwidth}
		\includegraphics[width=0.6\linewidth]{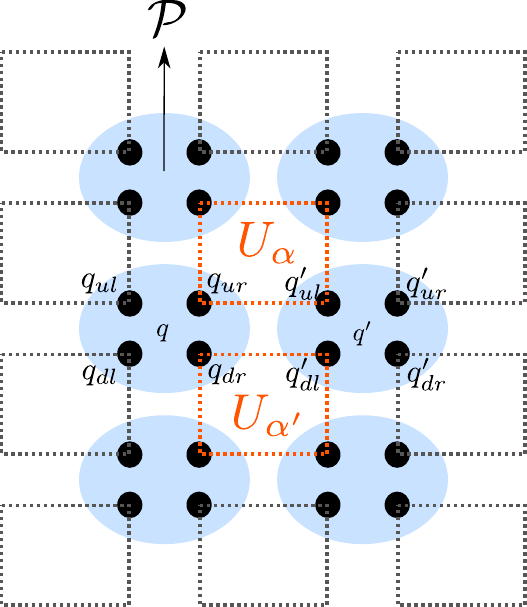}
		\caption{Bravy-type unitary operators.} \label{fig:b_directsum}
	\end{subfigure}%
	\caption{(a) Representation of a toric-code Hamiltonian, in which two types of plaquette operators alternate. (b) Representation of a state that is obtained by mapping four virtual particles (black dots) into a physical one (blue shape). The shaded lines represent unitary operators acting on adjacent virtual sites, each of them belonging to a different qudit.} \label{fig:direct_sum_toric}
\end{figure}

It follows that it suffices to find a set of pairwise commuting unitary operators in order to construct an operator $K \in \mathbb{K}_{r,M}$. We will explicitly outline three different methods on how to construct such a set.
\newline
\newline
\textit{Diagonal unitaries}.  
The first obvious way to do so is to take all the unitary operators diagonal on the same basis.  Note that one can use these unitary operators to construct graph states \cite{cluster, Hein_2004} (such as the cluster state \cite{cluster}) and weighted graph states \cite{hartmann2007weighted} by applying them over a lattice where each edge represents a maximally entangled state between adjacent vertices. Thus, the family of states \eqref{Psibetat} trivially contains those.
\newline
\newline
\textit{Toric-code-type unitaries}.  
An alternative approach for two-dimensional rectangular lattices is as follows.  First, take some Hermitian operators $h_A$ and $h_B$ acting on adjacent plaquettes $A$ and $B$ and assume that $h_{A,B}$ are of a Toric-code type \cite{Kitaev_2003} (see Figure \ref{fig:a_toric}). By this we mean that $h_A = \bigotimes_{i \in A} O_A$, $h_B = \bigotimes_{i \in B} O_B$, where $O_{A,B} = O_{A,B}^\dagger$ and such that $\left\{ O_A, O_B \right\} = 0$ (for the Toric code we have $O_A = \sigma_z, O_B = \sigma_x$). Since the supports of $h_A$ and $h_B$ intersect in two vertices, the anticommutativity condition implies that $\left[h_A, h_B \right] = 0$. Thus, we can generate commuting unitary operators by evolving $h_{A,B}$ up to different times:
\begin{equation}
\begin{split}
        U^A_{\alpha} = e^{i h_A {t_\alpha}}, \quad  U^B_{\beta} = e^{i h_B {t_{\beta}}}.
\end{split}
\label{eq_unitaries_toric}
\end{equation}
By construction, we have $\left[U^A_\alpha , U^A_{\alpha'} \right] = \left[U^A_\alpha , U^B_{\beta} \right] = \left[U^B_\beta , U^B_{\beta'} \right] = 0$ for all $\alpha,\alpha',\beta,\beta'$. Note that we can choose $O(N)$ free parameters, corresponding to the different evolution times $\lbrace t_\alpha \rbrace_\alpha$ and $ \lbrace t_\beta \rbrace_\beta$,  as many as plaquettes in the lattice.
\newline
\newline
\textit{Bravy-type unitaries}.  
Finally, we present a more general approach, which is a generalization of the results of Bravyi et al. \cite{Bravyi2005CommutativeVO} about the characterization of commuting, local Hamiltonians in $1$D (see also \cite{Aharonov2011OnTC} for a discussion on this topic). We will present the method for the particular case of a $2$D rectangular locality structure, but the generalization to other geometries is straightforward.

Let us start with a state defined over a lattice. At each site, we have four virtual particles of dimension $D$, which are mapped into a single physical qudit, with a physical degree of freedom $d$. For a visualization see Figure \ref{fig:b_directsum}, where we associate black dots with virtual particles; the big blue circles represent the physical sites, and correspond to the vertices of the lattice. We denote by $\mathcal{P}: (\mathbb{C}^{d})^{\otimes 4} \mapsto \mathbb{C}^D$ the map from the virtual to the physical degrees of freedom. We take $\mathcal{P}$ to be unitary, which means that $d^4 = D$. 

Let us denote by $\mathcal{H}_q \cong \mathbb{C}^d$ the Hilbert space associated with the physical particle at vertex $q$. Bravy and Vyalyi \cite{Bravyi2005CommutativeVO} showed that $\mathcal{H}_q$ can be decomposed as: 
\begin{equation}
       \mathcal{H}_q = \bigoplus_{i} \left( \mathcal{H}_{q_{ul}}^i \otimes \mathcal{H}_{q_{ur}}^i \otimes \mathcal{H}_{q_{dl}}^i  \otimes \mathcal{H}_{q_{dr}}^i \right),
\end{equation}
where $q_{ul},q_{ur},q_{dl},q_{dr}$ correspond to the virtual degrees of freedom, see Fig. \ref{fig:b_directsum}. We denote by $\mathcal{H}_{q_{ul}}$ the Hilbert space associated with the virtual particle $q_{ul}$ and by  $\mathcal{H}^i_{q_{ul}}$ a subspace of $\mathcal{H}_{q_{ul}}$.

Let us now define some unitary operators $U_\alpha$, acting on four adjacent virtual particles, each belonging to a different vertex as in Fig. \ref{fig:b_directsum}. Here $\alpha$ just denotes an index to enumerate the different operators, without any spatial meaning. We do not impose any restriction on these operators apart from the virtual particles in which they act. Let us now show that when we consider the operators $U_\alpha$ acting on the joint Hilbert space of two neighbouring physical sites, i.e. on $\mathcal{H}_q \otimes \mathcal{H}_{q'}$, they commute. To see this, consider the examples from Fig. \ref{fig:b_directsum}. We can write the operators $U_\alpha$ and $U_\alpha'$ as:
\begin{eqnarray}
 U_\alpha|_{\mathcal{H}_q \otimes \mathcal{H}_{q'} } =
o_{q_{ur}q'_{ul}} \otimes \mathds{1}_{q_{ul} q_{dr} q_{dl}} \otimes \mathds{1}_{q'_{ur} q'_{dl} q'_{dr}} \nonumber \\
 U_\alpha'|_{\mathcal{H}_q \otimes \mathcal{H}_{q'} } =
o_{q_{dr}q'_{dl}} \otimes \mathds{1}_{q_{ul} q_{ur} q_{dl}} \otimes \mathds{1}_{q'_{ur} q'_{ul} q'_{dr}}
\end{eqnarray}
for some operators $o_{q_{ur}q'_{ul}}$ and $o_{q_{dr}q'_{dl}}$ that act non-trivially only on the subspaces $\mathcal{H}_{q_{ur}} \otimes \mathcal{H}_{q'_{ul}}$ and $\mathcal{H}_{q_{dr}} \otimes \mathcal{H}_{q'_{dl}}$ respectively. With this decomposition, it is easy to see that $\left[U_\alpha, U_{\alpha'} \right] = 0$.

Now note that the application of the unitary operators $\mathcal{P}$ does not affect the commutation; it simply changes the basis in which the qudits are expressed. Hence we can define the final set of unitaries to be:
\be
\left( \prod_{\mathcal{H}_q \cap \lambda (U_\alpha ) \neq \emptyset}  \mathcal{P}_q^\dagger \right) U_\alpha \left( \prod_{\mathcal{H}_q \cap \lambda (U_\alpha ) \neq \emptyset} \mathcal{P}_q \right),
\ee
where $\lambda(U_\alpha)$ denotes as before the support of the operator $U_\alpha$. Note that this defines a set of unitary operators over the full lattice, such that all of the operators commute pairwise.

Note that this technique does not cover the generation of states where $d=2$. Therefore for qubits other approaches like the ones presented above must be employed.

\section{Continuity bound on the gap} \label{appendix_continuity_gap}

Here, we show that the validity of condition \eqref{hnm}
for $h_i$, Eq.~\ref{hj}, at a given point $H(\tau,t)$ (with $\tau$ the imaginary time formerly known as $\beta$), implies the validity of the same condition at a different point $(\tau',t)$, with $\Delta\tau=\tau'-\tau>0$, with Hamiltonian terms $h_i'$, only with different values $c'_{ij}$ and $c_{ji}'$. 
Recall \eqref{hnm}:
\begin{equation}
\
 \forall i\ne j:\,
  0\preceq h_i h_j + h_j h_i + a_{ij} h_i^2 + a_{ji} h_j^2 - c_{ij}h_i - c_{ji} h_j\ .
  \label{app:hnm}
\end{equation}
Let us pick a specific pair $i\ne j$. Let
\begin{equation}
A=\prod_{m\in {\nu_i\setminus\nu_j}}e^{-\Delta\tau\kappa_{1,m}}\ ,\quad
C=\prod_{m\in {\nu_j\setminus\nu_i}}e^{-\Delta\tau\kappa_{1,m}}\ ,\quad
B=\prod_{m\in \nu_i\cap\nu_j}e^{-\Delta\tau\kappa_{1,m}}
\end{equation}
be the part of the deformation $e^{(\Delta\tau) K_1}$ acting only on $h_i$, only $h_j$, and both $h_i$ and $h_j$, respectively, that is,
\begin{equation}
h_i' = AB h_i AB\ ,\quad h_j'=BC h_j BC\ .
\end{equation}
By construction,
\begin{equation}\label{app-ABC}
\alpha\openone\le A^2\le \openone\ ,\quad
\beta\openone\le B^2 \le\openone\ ,\quad
\gamma\openone\le C^2 \le\openone\ ,
\end{equation}
where (unless we have tighter bounds)
\begin{equation}
\label{app-sdp-alpha-beta-gamma}
\alpha=e^{-2|\nu_i\setminus\nu_j|\Delta\tau}\ ,\quad
\gamma=e^{-2|\nu_j\setminus\nu_i|\Delta\tau}\ ,\quad
\beta=e^{-2|\nu_i\cap\nu_j|\Delta\tau}
\end{equation}
(in particular, $\alpha,\beta,\gamma>0$).
Henceforth, for convenience of notation we will write $P:=h_i$ and $Q:=h_j$, as well as  $a:=a_{ij}$, $b:=a_{ji}$,  $c:=c_{ij}$, $d=c_{ji}$.

Let us now derive two inequalities. First,
\begin{align}
\beta(PQ+QP)
&=
\beta(P+Q)^2-\beta P^2-\beta Q^2
\\
&\le
(P+Q)B^2(P+Q)-\beta P^2 -\beta Q^2
\\
\label{gap-lemma1-line3}
&=
PB^2Q + QB^2P + PB^2P-\beta P^2 + QB^2Q-\beta Q^2\ .
\end{align}
Second, using that $\openone\le C^{-2}$, $-\openone\le -\gamma C^{-2}$, 
$-A^2\le\alpha\openone$, and $B^2\le\openone$,
we have that -- for now assuming $0\le a\le 1$ -- 
\begin{align}
PB^2P-\beta(1-a)P^2-\beta c P
&  =
 aPA^2B^2P\otimes C^{-2}-
aPA^2B^2P\otimes C^{-2}+
PB^2P
- \beta(1-a)P^2-\beta c P
\\
&\le
aPA^2B^2P\otimes C^{-2}+
\left[(1-a\alpha) PB^2P  
- \gamma\beta(1-a)P^2-\gamma\beta c P\right]\otimes C^{-2}
\\
\label{gap-lemma2-line2}
& \le
aPA^2B^2P\otimes C^{-2}+\big\{   \underbrace{[(1-a\alpha)-\gamma\beta(1-a)]}_{=:X} P^2-\gamma\beta c P
\big\}\otimes C^{-2}
\\
&  \leq
\label{gap-lemma2-line4}
aPA^2B^2P\otimes C^{-2}-c' P\otimes C^{-2}\ ,
\end{align}
where we have used that the term labelled $X$ in \eqref{gap-lemma2-line2} is non-negative (as $(1-a\alpha)>(1-a)$ and $\gamma\beta<1$) to bound $XP^2\le XP$, and defined
\begin{subequations}
\label{app-sdp-cprime}
\begin{align}
&& c' & :=
\gamma\beta c - (1-\gamma\beta) - a(\gamma\beta-\alpha) 
 && \mbox{for\ }0\le a \le 1\ .&&\qquad\qquad\\
\intertext{The same inequality can be shown to hold in the other cases, with}
&& c'&:=\gamma\beta c - (1-\gamma\beta) - a(\gamma\beta-1) && \mbox{for\ }a<0\ ;
\\
&& c'&:=\gamma\beta c - (1-\beta) - a(\beta-\alpha) && \mbox{for\ } a>1, 1-a\alpha\ge 0 \ ;
\\
&&c'&:=\gamma\beta c - (1-\beta) - a\beta(1-\alpha) && \mbox{for\ } 1-a\alpha< 0 \ .
\end{align}
\end{subequations}

The same way, we can also obtain for $0\le b\le 1$
\begin{equation}
\label{gap-lemma3}
QB^2Q-\beta(1-b)Q^2-\beta d Q
\le
bA^{-2}\otimes QB^2C^2Q - d' A^{-2}\otimes Q\ ,
\end{equation}
where
%$d'$ is obtained from $c'$ by exchanging $\alpha$ and $\gamma$, and replacing $c$ by $d$.
\begin{equation}
\label{app-sdp-dprime}
d' :=
\alpha\beta d - (1-\alpha\beta) - b(\alpha\beta-\gamma)
\end{equation}
(or correspondingly -- with $\alpha$ and $\gamma$ exchanged w.r.t.\ $c'$ -- when $b<0$ or $b>1$).
Starting from \eqref{app:hnm}, we thus obtain
\begin{align}
0
&\stackrel{\eqref{app:hnm}}{\le}
\beta (PQ+QP+aP^2+bQ^2-cP-dQ)
\\
&\stackrel{\eqref{gap-lemma1-line3}}{\le}
(PB^2Q + QB^2P +  PB^2P-\beta P^2 + QB^2Q-\beta Q^2) +\beta(aP^2+bQ^2-cP-dQ) 
\\
&\stackrel{(\ref{gap-lemma2-line4},\ref{gap-lemma3})}{\le}
PB^2Q + QB^2P +aPA^2B^2P\otimes C^{-2}
bA^{-2}\otimes QB^2C^2Q - c' P\otimes C^{-2} - d' A^{-2}\otimes Q\ .
\end{align}
Conjugating both sides of the equation with $ABC$, we now immediately see that this is nothing but the condition \eqref{app:hnm} for the Hamiltonian $h_k'$, with $a_{ij}$ and $a_{ji}$ unaltered, and new feasible points $c_{ij}'$ and $c_{ji}'$ as given by Eqs.~\eqref{app-sdp-cprime} and \eqref{app-sdp-dprime}. 

Importantly, in case $0\le a_{ij}\le 1$ for all $i,j$,
using \eqref{app-sdp-alpha-beta-gamma} this yields a feasible point with
\begin{equation}
c_{ij}' =  e^{-2|\nu_j|\Delta\tau} c_{ij} - (1-e^{-2|\nu_j|\Delta\tau}) - a_{ij}\big(e^{-2|\nu_j|\Delta\tau}-
e^{-2|\nu_i{\setminus}\nu_j|\Delta\tau}\big) \ ,
\end{equation}
for \emph{all} $c_{ij}$ in 
\eqref{hnm}, which immediately 
implies a continuity bound on the gap which only depends on the geometry of the model,
but not on the lattice size or the specifics of the model at hand. This can be straightforwardly adapted to the case where some $a_{ij}$ are negative or above $1$.

As outlined in the main text, this bound can subsequently used, starting from $\beta=0$, to determine a regime for which the gap can be lower bounded by a uniform $\Delta_0>0$. Clearly, tighter bounds can be obtained by choosing more intermediate values of $\beta$.

\section{Symmetrization of the SDP for translational invariant systems}\label{appendix_symmetrization}

Let us see that for translational-invariant systems, the SDP presented in the main text gives a bound $\delta$ which is independent of the system size. For simplicity, we consider 1D systems and the case in which Eq. \eqref{sdp-only-overlapping} holds. The most general case, as well as the higher dimensional case, can be generalized from this proof. 

Let $N$ denote the system size and let us consider periodic boundary conditions. Consider the SDP \eqref{sdp}. Let us assume that we have a feasible point for some choice of the variables $(a^*_{ij}, c^*_{ij})$. Let us show that we can redefine these variables, such as we still have a feasible point with the same value of $\delta$, as follows: 
\begin{align}
 \tilde{a}_{i,i+1}:= \frac{1}{N} \sum_{i} a^*_{i,i+1}; \qquad  \tilde{a}_{i,i-1}:= \frac{1}{N} \sum_{i} a^*_{i,i-1} .\\
 \tilde{c}_{i,i+1} := \frac{1}{N} \sum_{i} c^*_{i,i+1}; \qquad 
\tilde{c}_{i,i-1} := \frac{1}{N} \sum_{i} c^*_{i,i-1} .
\end{align}
It is easy to see that the constraints \eqref{hnm} are still satisfied. Indeed: 
\begin{align*}
&h_{i} h_{i+1}+h_{i+1} h_{i}+\tilde{a}_{i,i+1} h_{i}^{2}+\tilde{a}_{i+1,i} h_{i+1}^{2}-\tilde{c}_{1} h_{i}-\tilde{c}_{2} h_{i+1} = \frac{1}{N} \left[ N (h_{i} h_{i+1}+h_{i+1} h_{i})   +\left( \sum_{k=1}^N a^*_{k,k+1} \right) 
h_i^2 + \right.\\
&+ \left. \left( \sum_{k=1}^N a^*_{k+1,k} \right)h_{i+1}^2 -   \left( \sum_{k=1}^N c^*_{k,k+1} \right)h_i - \left( \sum_{k=1}^N c^*_{k+1,k} \right)h_{i+1}   \right].
\end{align*}
Now note that, because of the translational invariance, we have that the terms $h_i h_{i+1}$, $h_{i+1}h_i $, $h_i = h_i \otimes \mathds{1}$ and $\mathds{1} \otimes h_{i+1}$ are the same for all $i$. Thus, we can write $N (h_i h_{i+1} +h_{i+1}h_i) = \sum_{k=1}^N h_k h_{k+1} +h_{k+1}h_k$, and get:
 \begin{align*}
 	&h_{i} h_{i+1}+h_{i+1} h_{i}+\tilde{a}_{i,i+1} h_{i}^{2}+\tilde{a}_{i+1,i} h_{i+1}^{2}-\tilde{c}_{i,i+1} h_{i}-\tilde{c}_{i+1,i} h_{i+1} = \\
 	=& \frac{1}{N} \sum_{k=1}^N \left[\underbrace{h_{k} h_{k+1}+h_{k+1} h_{k}+{a}^*_{k,k+1} h_{k}^{2}+{a}^*_{k+1,k} h_{k+1}^{2}-{c}^*_{k,k+1} h_{k}-{c}^*_{k+1,k} h_{k+1} }_{\succeq 0} \right]  \succeq 0.
 \end{align*}
For the constraints \eqref{sdp3}-\eqref{sdp4} we get:
\begin{align*}
\sum_{j\neq i} \tilde{a}_{ij} = &  \tilde{a}_{i,i+1} +  \tilde{a}_{i,i-1} =  \frac{1}{N}  \sum_{i=1}^N  a^*_{i,i+1} +  \frac{1}{N}  \sum_{i=1}^N  a^*_{i,i-1} =   \frac{1}{N} \sum_{i=1}^N  \left(\underbrace{a^*_{i,i+1} +  a^*_{i,i-1} }_{=1}\right) = 1. \\
\sum_{j\neq i} \tilde{c}_{ij} = &  \tilde{c}_{i,i+1} +  \tilde{c}_{i,i-1} =  \frac{1}{N}  \sum_{i=1}^N  c^*_{i,i+1} +  \frac{1}{N}  \sum_{i=1}^N  c^*_{i,i-1} =   \frac{1}{N} \sum_{i=1}^N  \left(\underbrace{c^*_{i,i+1} +  c^*_{i,i-1} }_{=x}\right) = x. 
\end{align*}
With the new variables $\tilde{a}_{ij}, \tilde{c}_{ij}$ we actually have redundant information in \eqref{hnm} and it suffices to write down a single condition: 
\begin{align}\label{SDP-new}
h_{i} h_{i+1}+h_{i+1} h_{i}+\tilde{a}_{i,i+1} h_{i}^{2}+\tilde{a}_{i+1,i} h_{i+1}^{2}-\tilde{c}_{i,i+1} h_{i}-\tilde{c}_{i+1,i} h_{i+1}  \succeq 0\\
\tilde{a}_{i,i+1}+\tilde{a}_{i, i-1} = 1 \\
\tilde{c}_{i,i+1}+\tilde{c}_{i, i-1} = x
\end{align}
It is easy to see now that adding a new particle, i.e., $N \rightarrow N+1$, does not change the SDP (since it will only add redundant information to  \eqref{SDP-new}). Thus, given that for a TI  finite system our method provides a non-trivial bound for the gap, the same bound holds for $N \rightarrow \infty$. In this case, we recover the result of the martingale method by Ref. \cite{Werner}. Note that for higher dimension, instead of a single condition \eqref{SDP-new} we would have as many as different types of overlaps within the different Hamiltonian terms.

\section{Bound on the norms of the observables $Q_\lambda$}\label{appendix_norms_Olambda}

Let us consider the operators $Q_\lambda$ be defined as in Eq. \eqref{eq:q-lambda}, i.e. $Q_\lambda = O_\lambda^\dagger P O_\lambda$, with $O_\lambda$ as defined in Eq. \eqref{eq:o-lambda} and $P$ being a Pauli string. We consider the infinite-norm $\Vert Q_\lambda \Vert_\infty$ and from now on we will drop the subindex for simplicity. Let us write the following inequality for the norm of a product of operators:  for $A$ and $B$ being invertible operators, it holds that:
\begin{equation} \label{AB_ineq}
\Vert A B \Vert  \geq \Vert A \Vert \frac{1}{\Vert B^{-1} \Vert}.
\end{equation}
Eq. \eqref{AB_ineq} can be proven by using the submultiplicity of the norm, namely:
\begin{equation}
\Vert A \Vert = \Vert A  B B^{-1}\Vert \leq \Vert  A  B \Vert \Vert B^{-1}\ \Vert .
\end{equation}
Let us first bound the norm of $O_\lambda$ and $ \left( O_\lambda^\dagger \right)$. With this and by means of Eq. \eqref{AB_ineq}, we will be able to lower bound $\Vert Q_\lambda \Vert$.

Let us take the norm of $O_\lambda$ and insert its definition from Eq. \eqref{eq:o-lambda}:
\begin{align*}
	\Vert O_\lambda \Vert = &\Vert   \prod_{n \in \mu (\lambda)} e^{-i \kappa_{2,n}} \prod_{m \in \nu (\lambda)}   e^{-\beta \kappa_{1,m}} \Vert   \underbrace{\geq}_{Eq. \eqref{AB_ineq}} \underbrace{\Vert  \prod_{n \in \mu (\lambda)} e^{-i \kappa_{2,n}}  \Vert }_{=1} \prod_{m \in \nu (\lambda)} \Vert  e^{\beta \kappa_{1,m}} \Vert^{-1} =   \prod_{m \in \nu (\lambda)} \Vert  e^{\beta \kappa_{1,m}} \Vert^{-1} \geq  \\
	& \geq \prod_{m \in \nu (\lambda)} e^{-\beta} = e^{-\beta |\nu(\lambda)|}.
\end{align*}
Here we have used that $\left[ \kappa_{1,r}, \kappa_{1,s}\right]= 0$, so we can just write the inverse $\left(  \prod_{m \in \nu (\lambda)}   e^{-\beta \kappa_{1,m}} \right)^{-1} =  \prod_{m \in \nu (\lambda)}  \left(  e^{-\beta \kappa_{1,m}} \right)^{-1}.$ We can now use that $\left( e^{-\beta \kappa_{1,m}}  \right)^{-1} =  e^{\beta \kappa_{1,m}}$ and that $\Vert e^{\beta \kappa_{1,m}} \Vert_\infty \leq  e^{\beta}$. With this, it follows that: 
\begin{equation} \label{bound_O}
\Vert O_\lambda \Vert = \Omega( e^{-\beta |\nu_\lambda|}).
\end{equation}

Let us now study$ \left( O_\lambda^\dagger \right)$:
\begin{align*}
	\Vert \left( O_\lambda^\dagger  \right)^{-1} \Vert  =& \Vert \left( \prod_{m \in \nu(\lambda)} e^{-\beta \kappa_{1,m}}  \prod_{n \in \mu(\lambda)} e^{i \kappa_{2,n}} \right)^{-1}\Vert  = \Vert \prod_{n \in \mu(\lambda)} e^{-i \kappa_{2,n}} \prod_{m \in \nu(\lambda)} e^{\beta \kappa_{1,m}}  \Vert \underbrace{\leq}_{\text{subm.}} \underbrace{ \Vert \prod_{n \in \mu(\lambda)} e^{-i \kappa_{2,n}} \Vert}_{=1} \prod_{m \in \nu(\lambda)}  \Vert e^{\beta \kappa_{1,m}}  \Vert \leq   \\
	\leq & \prod_{m \in \nu(\lambda)} e^{\beta} = e^{\beta |\nu(\lambda)|}.
	\end{align*}
The derivation above implies the following bound:
\begin{equation} \label{bound_O-1}
	\frac{1}{	\Vert \left( O_\lambda^\dagger  \right)^{-1} \Vert }  \geq e^{-\beta |\nu(\lambda)|} \ .
\end{equation}

Note that in all the bounds above it is present the cardinality of $| \nu(\lambda)|$, so let us estimate this value in terms of the size $|\lambda|$: 
\begin{equation} \label{cardinality_nu}
	| \nu (\lambda)| = |\underset{j\in \lambda}{\bigcup \nu_j}| \leq \sum_{j\in \lambda} |\nu_j | = \mathcal{O} (|\lambda| z^{2r_1}).
\end{equation}
Where we have used that $|\nu_j | = \mathcal{O}(z^{2r_1})$, where $z$ is the degree of incidence of the graph at which vertex the particles are placed and $r_1$ is the radius of the operators $\kappa_1$.

With all the derivations above, now we can get a lower bound on $\Vert Q_\lambda \Vert$:
\begin{align*}
\Vert Q_\lambda \Vert = \Vert O_\lambda  P O_\lambda^\dagger \Vert  \underbrace{\geq}_{Eq. \eqref{AB_ineq}} \Vert O_\lambda \Vert \Vert \left( P O_\lambda^\dagger \right)^{-1}  \Vert^{-1} = \Vert O_\lambda \Vert  \cdot \frac{1}{\Vert \left( O_\lambda^\dagger \right)^{-1} P \Vert }.
\end{align*}
where we have use that $P^{-1}= P$. Now we can use submultiplicity of the norm and Eq. \eqref{bound_O-1} to get:
\begin{equation}
 \frac{1}{\Vert \left( O_\lambda^\dagger \right)^{-1} P \Vert } \geq \frac{1}{\Vert \left( O_\lambda^\dagger \right)^{-1} \Vert } \underbrace{\frac{1}{\Vert P \Vert}}_{=1} \geq e^{-\beta |\nu(\lambda)|}
\end{equation} 

With this and Eq.  \eqref{bound_O} and Eq. \eqref{cardinality_nu} now we get: 
\begin{equation}
\boxed{\Vert Q_\lambda \Vert  \geq e^{-2\beta |\nu(\lambda)|} = \Omega (e^{-2\beta |\lambda| z^{2r_1}})}
\end{equation}

This means that $\Vert Q_\lambda \Vert $ decays \emph{at most} as $\mathcal{O}(e^{-2\beta |\lambda| z^{2r_1}})$. If we consider now the terms such that $|\lambda| $ is constant (or at most goes as $\mathcal{O}(\log N)$, with $N$ being the system size) then we get that $\Vert Q_\lambda \Vert $ decays at most polynomially with the system size.

\section{Analysis of the verification test}
\label{appendix_bounds_verification}
In this section, we analyze the verification test for the setup presented in Section \ref{section_quantum_verification}, that is, in which the verifier can perform local measurements to the state that the prover  has prepared. This case is analogous to the one presented in \cite{Hangleiter_2017}. We modify it here by imposing further restrictions on the prover: we assume that that he can only perform Pauli measurements, which might be more suitable for experiments carried out with NISQ devices.  

We first analyze the probability that, for a given tolerance threshold $\epsilon$, the empirical mean $\langle h_j \rangle_P$ computed through the measurements (where the subindex $P$ emphasizes that the verifier measures the state that the prover has prepared) is above such threshold, due to statistical noise produced by finite sampling. This statistical fluctuation depends on the variance of the distribution $\langle h_j \rangle_P$, which is computed by decomposing the operators in the Pauli basis. This variance is bounded in Appendix \ref{appendix_bound_variance}.

\subsection{Bounds on the fidelity and the number of measurement rounds}
Let us denote by $\langle h_j \rangle_P$ the random variable result of estimating the expectation value of $h_j$ by measuring the state in the Pauli basis, for a number of samples $L_j$. We define this random variable as $\langle h_j \rangle_P = \frac{1}{L_j} \sum_{k=1}^{L_j} \hvar_j^{(k)}$, where $\hvar_j$ refers to the random variable associated to $\langle h_j \rangle_P$ computed with a single round (see next section and Eq. \eqref{eq_mathcalh_exp_value} for a formal definition of $\hvar$). As before, we assume that $L_j = L_{j'}$ and that we do not recycle samples, i.e., we always use different samples for different local terms $h_j$. In the same way, we denote by $\langle H \rangle_P$ the estimator of the energy, which is a random variable whose mean we denote by $\mathbb{E}\left[ \langle H \rangle_P\right]$. We also define $F_P= 1 - \frac{\langle H \rangle_P}{\delta}$, with $\delta$ being the lower bound on the spectral gap computed in Section \ref{section_GapsCorr}. By Eq. \eqref{bound} $F_P$ is an estimator of a lower bound on the fidelity between the prover's state and the true ground state. Then, we get the following result: for $\alpha \in (0,1)$ and a threshold parameter $\epsilon>0$, it holds that
\begin{equation} \label{eq:bound_fidelity}
    P \left(F_P < \mathbb{E} \left[ F_P \right] - \epsilon \right) \leq \alpha,
\end{equation}
as long as the number of samples $L_j$ satisfies: 
\begin{equation} \label{eq:bound_Mn}
    L_j \geq \frac{2^{5 |\lambda| + 1}}{\delta^2 \epsilon^2} \log\left(\frac{1}{\alpha^{1/N}}\right),
\end{equation}
where $N$ is the number of terms in the Hamiltonian and $|\lambda|$ is its locality.

\emph{Proof}. By Eq. \eqref{bound} we get that: 
\begin{equation} \label{eq:prob_H}
     P \left(F_P < \mathbb{E} \left[ F_P \right] - \epsilon \right) = P \left( \langle H \rangle_P > \mathbb{E} \left[ \langle H \rangle_P \right] + \delta \epsilon \right).
\end{equation}
Recall that $\langle H \rangle_P = \sum_j \langle h_j \rangle_P$. We can assume that the numbers of samples used to estimate any $\langle h_j \rangle_P$ is large enough so the central limit theorem applies. Thus, we can assume that $\langle h_j \rangle_P \sim \mathcal{N}\left( \mathbb{E}[\langle h_j \rangle_P], \mathrm{Var}[\langle h_j \rangle_P]\right) $ follows a Gaussian distribution, which therefore implies that  $\langle H \rangle_P$ is Gaussian itself. Then, we can bound \eqref{eq:prob_H} by: 
\begin{equation}
    P \left( \langle H \rangle_P > \mathbb{E} \left[ \langle H \rangle_P \right] + \delta \epsilon \right) \leq e^{- (\delta \epsilon)^2/2 \mathrm{Var}[\langle H \rangle_P]} = \prod_j e^{- (\delta \epsilon)^2/2 \mathrm{Var}[\langle h_j \rangle_P]},
\end{equation}
where the inequality follows from bounding the tails oft he Gaussian distribution and the last equality comes from the fact that $\mathrm{Var}[\langle H \rangle_P] = \sum_j \mathrm{Var}[\langle h_j \rangle_P]$ since the random variables $\langle h \rangle_P$ are independent. Note that $\mathrm{Var}[\langle h_j \rangle_P ] = \mathrm{Var}[\hvar_j]/L_j$. We will see in the following section that: 
\begin{equation}
    \mathrm{Var}[\hvar_j] \leq 2^{5|\lambda|}.
\end{equation}
By substituting this into the previous equation and imposing the bound \eqref{eq:bound_Mn}  on $L_j$ we get: 
\begin{equation}
    P \left( \langle H \rangle_P > \mathbb{E} \left[ \langle H \rangle_P \right] + \delta \epsilon \right)  \leq \alpha,
\end{equation}
which thus proves Eq. \eqref{eq:bound_fidelity}. Note that this equation is similar to Eq. (7) and (15) in \cite{Hangleiter_2017} and a derivation in terms of Hoeffding bounds as in this paper would also apply here --- although the difference is again that we do not measure the eigenenergies of $h$ directly but only the expectation values of strings of Pauli operators.

\subsection{Bound on the variance of $\langle h \rangle_P$}
\label{appendix_bound_variance}

Let us now study the properties of the distribution of $\langle h \rangle_P$ for a fixed local term $h$.  We first introduce some preliminary definitions, some of which were already introduced in the main text, that will help in the upcoming discussion:
\begin{itemize}
	\item We define a set random variables $\alpha_j$, for $j = 1,...,N$. Each variable  $\alpha_j$  can take the values $\lbrace x, y, z \rbrace$ with equal probability. We denote the set of this variables by $B$, $B = \lbrace \alpha_j \rbrace_{j=1}^N$. Note that each element of  $B$ corresponds to the basis in which we choose to measure each qubit independently.
	
	\item We denote by  $J$ any of the subsets of indices of $ \lambda(h) = \lbrace j_1,...,j_{|\lambda|} \rbrace$. Note that the cardinal of $J$ goes from $0,...,{|\lambda|}$, where we also include the empty set for convenience. In the same fashion, we denote by $B_J$ the subset of random variables $\alpha_j$ as defined before with corresponding indices $j \in J$.
	\item Let $\randsigma^J_{B_J} = \randsigma^{j_1}_{\alpha_{j_1}}\cdots \randsigma^{j_{\vert J \vert }}_{\alpha_{j_{\vert J \vert}}} $ be the random variable with possible outputs $\pm 1$ resulting from measuring the qubits $j \in J$ in the basis $\alpha_{j} \in B_J$.
\end{itemize}

With these preliminaries we can now define the following random variable $\hvar$:
\begin{equation}
	\begin{split}
		\hvar & = o_0 + \sum_{j=1}^{|\lambda|} 3 \cdot o_{\alpha_j}^{j} \randsigma_{\alpha_j}^j + \sum_{j,k = 1}^{|\lambda|} 3^2 \cdot o_{\alpha_j, \alpha_k}^{j,k} \randsigma_{\alpha_j}^j \randsigma_{\alpha_k}^k + ... + 3^{|\lambda|} \cdot o_{\alpha_{j_1} \alpha_{j_2},..., \alpha_{j_{|\lambda|}}}^{j_1,...,j_{|\lambda|}} \randsigma_{\alpha_{j_1}}^{j_1}... \randsigma_{\alpha_{j_{|\lambda|}}}^{j_{|\lambda|}} =  \sum_{J \subseteq \lambda(h)}^{|\lambda|} 3^{\vert J \vert }o_{B_J}^J \randsigma_{B_J}^J.
	\end{split}
	\label{eq_h_single_round}
\end{equation}
 Here the coefficients $o_{\alpha_{j_1} \alpha_{j_2}... \alpha_{j_{|\lambda|}}}^{j_1...j_{|\lambda|}}$ are those of the Pauli expansion of  $\langle h \rangle$:
\begin{equation}
	\begin{split}
		\langle h \rangle&= \sum_{j=1}^n \sum_{\gamma = x, y, z} o_{\gamma}^j \langle  \paulisigma^j_\gamma \rangle + \sum_{j,k=1}^n \sum_{\gamma_1, \gamma_2 = x,y,z} o_{\gamma_1 \gamma_2}^{j k} \langle \paulisigma^j_{\gamma_1} \paulisigma^k_{\gamma_2} \rangle 
		+  \ldots +  \sum_{j_1, \ldots,j_k= 1}^n\sum_{\gamma_{j_1},\ldots,\gamma_{j_k}= x,y,z} o^{j_1\ldots j_{|\lambda|}}_{\gamma_{j_1}\ldots\gamma_{j_{|\lambda|}}} \langle \paulisigma_{\gamma_{j_1}}^{j_1} \cdots \paulisigma_{\gamma_{j_k}}^{j_k}  \rangle.
	\end{split}
	\label{eq_h_pauli_exp}
\end{equation}
Note that in the definition of the variable $\hvar$ in Eq. \eqref{eq_h_single_round} we do not sum over all possible basis choices: this variable actually corresponds to the computation of $\langle h\rangle_P$ with a single round, in which we only measure the state once in some basis $\bm\alpha = (\alpha_1,...,\alpha_N)$. Note that this is different from Eq. \eqref{eq:O-sbar}, in which we averaged over different rounds with the same basis choice (one could in principle also perform an analysis with multi-rounds, but then $\mathcal{R}(\bm \gamma)$ is itself a random variable, which makes the analysis more cumbersome).  With this single sample, we can only give an estimator for the following marginals
\begin{equation}
	\langle \paulisigma_{\alpha_j}^j \rangle, \langle \paulisigma^{j_1}_{\alpha_{j_1}} \paulisigma^{j_2}_{\alpha_{j_2}} \rangle , \ldots, \langle \paulisigma^{j_1}_{\alpha_{j_1}} \cdots \paulisigma^{j_{|\lambda|}}_{\alpha_{j_{|\lambda|}}}  \rangle,
\end{equation}
where $\alpha_{j_k}  \in B$. Let us now compute the expectation value of $\hvar$ :
\begin{equation} 
	\begin{split}
		\mathbb{E} \left[ \hvar\right] & = o_0 + \sum_{j=1}^{|\lambda|} \sum_{\gamma_j=x,y,z} 3 \cdot P(\alpha_j = \gamma_j) \mathbb{E} \left[  o_{\alpha_j}^{j} \randsigma_{\alpha_j}^j \right] +  \sum_{j,k = 1}^{|\lambda|} \sum_{\gamma_j,\gamma_k = x,y,z} 3^2 \cdot P(\alpha_j = \gamma_j, \alpha_k = \gamma_k) \mathbb{E} \left[ o_{\alpha_j, \alpha_k}^{j} \randsigma_{\alpha_j}^j \randsigma_{\alpha_k}^k \right] + \\
		& + \ldots +   \sum_{\gamma_{j_1},..., \gamma_{j_{|\lambda|}}} 3^{|\lambda|} \cdot P(\alpha_{j_1} = \gamma_{j_1},..., \alpha_{j_{|\lambda|}} = \gamma_{j_{|\lambda|}}) \mathbb{E} \left[ a_{\alpha_{j_1} \alpha_{j_2}... \alpha_{j_{|\lambda|}}}^{j_1...j_{|\lambda|}} \randsigma_{\alpha_{j_1}}^{j_1}... \randsigma_{\alpha_{j_{|\lambda|}}}^{j_{|\lambda|}} \right] = \\
		& = \sum_{J \subseteq \lambda(h)} \sum_{\gamma_1,...,\gamma_{|\lambda|} = x,y,z} 3^{\vert J \vert } P \left( B_J = ( \gamma_{j_1},...,\gamma_{j_{|\lambda|}} ) \right) o_{B_J}^J \mathbb{E} \left[\randsigma_{B_J}^J \right].
	\end{split}
	\label{eq_mathcalh_exp_value}
\end{equation}
Since $\alpha_j$ can take the values $\lbrace x,y,z \rbrace$ with equal probability we have that $P(\alpha_j = \gamma_j) = \displaystyle  \frac{1}{3}$, $\forall j, \forall \gamma_j$. Moreover, the variables $\alpha_j$ are independent and therefore the joint probability can be factorized:
$$P(\alpha_{j_1}= \gamma_{j_1},...,\alpha_{j_k}= \gamma_{j_k}) = P(\alpha_{j_1}= \gamma_{j_1}) \cdots P(\alpha_{j_k}= \gamma_{j_k}) = \left( \displaystyle \frac{1}{3} \right)^k, $$
for $k = 1,...,{|\lambda|}$. Note that Eq. \eqref{eq_mathcalh_exp_value} leads to the same expression as Eq.\eqref{eq_h_pauli_exp} and therefore:
\begin{equation}
	\langle h \rangle = \mathbb{E} \left[ \hvar\right]
	\label{eq_exp_values_h_h}.
\end{equation}

In statistical terms, $\hvar$ is an unbiased estimator for $\langle h \rangle$. We can now compute the variance of the variable $\hvar$ .  Now, note that $\randsigma^J_{B_J}$ and $\randsigma^{J'}_{B_{J'}}$ are not independent, since $J$ and $J'$ may have non-empty intersection. Therefore we express the variance of $\hvar$, defined by the sum in Eq. \eqref{eq_h_single_round}, in terms of the covariance of its terms:
\begin{align}
	\text{Var} \left[ 	\hvar \right] = \sum_{\vert J \vert, \vert J'\vert =0}^{|\lambda|} \ \sum_{J,J' \subseteq \lambda(h)} 3^{\vert J \vert} 3^{\vert J' \vert} a_{B_J}^J a_{B_J'}^{J'}  \text{Cov} \left[\randsigma^J_{B_J},\randsigma^{J'}_{B_{J'}} \right] .
	\label{eq_cov_alphas}
\end{align}

We can bound the coefficients $o_{B_J}^J o_{B_J'}^{J'}$ by $\left(o_{\text{max}}\right)^2$, where $o^2_{\text{max}} = \text{max} \left\{ \left(o_{B_J}^J \right)^2 \right\}$ and take it out of the sum for simplicity. Note that the covariance that appear in \eqref{eq_cov_alphas} can be bounded by:
\begin{equation}
	\text{Cov} \left[\randsigma^J_{B_J},\randsigma^{J'}_{B_{J'}} \right] \leq 1.
\end{equation}
Then:
\begin{equation}
	\begin{split}
		\text{Var} \left[ 	\hvar \right] & \leq \left(o_{\text{max}}\right)^2  \sum_{\vert J \vert, \vert J'\vert =0}^{|\lambda|} 3^{\vert J \vert} 3^{\vert J' \vert} \sum_{J,J' \subseteq \lambda(h)}  1 = \left(o_{\text{max}}\right)^2   \left( \sum_{\vert J \vert =1}^{|\lambda|} 3^{\vert J \vert}  \sum_{J \subseteq \lambda(h)} 1 \right)^2 =\\
		& =  \left(o_{\text{max}}\right)^2 \underbrace{ \left(\sum_{\vert J \vert =1}^{|\lambda|} \begin{pmatrix}
				{|\lambda|} \\ \vert J \vert
			\end{pmatrix}  3^{\vert J \vert}\right)^2 }_{4^{|\lambda|}} =  o_{\text{max}}^2 4^{2{|\lambda|}}.
	\end{split}
	\label{eq_bound_var_mathcalh}
\end{equation}
Note that $o^2_{\text{max}}$ depends on the particular local term that we are considering. We can get rid of this dependence by bounding it by the following value:
\begin{equation}
	o_{\text{max}}^2 \leq 2^{|\lambda|},
	\label{eq_amax_bound}
\end{equation}
which follows from the fact that $\text{tr} \left(h^2 \right) \leq 2^{|\lambda|}$.

\end{document}